%% file: mnras_dds_magnetic_env.tex
\documentclass[a4paper,fleqn,usenatbib]{mnras}

\usepackage{newtxtext,newtxmath}

\usepackage[T1]{fontenc}
\usepackage{ae,aecompl}

\usepackage{graphicx}	
\usepackage{amsmath}	
\usepackage{amssymb}	
\usepackage{pdflscape}	

\title[LPF Magnetic environment]{Spacecraft and interplanetary contributions to the magnetic environment on-board LISA Pathfinder}

\input{include/author_list_extended_mnras}

\date{Accepted 2020 March 18. Received 2020 March 15; in original form 2020 January 28}

\pubyear{2020}

\begin{document}
\label{firstpage}
\pagerange{\pageref{firstpage}--\pageref{lastpage}}
\maketitle

\begin{abstract}
LISA Pathfinder (LPF) has been a space-based mission designed to test new technologies that will be required for a gravitational wave observatory in space. Magnetically driven forces play a key role in the instrument sensitivity in the low-frequency regime (mHz and below), the measurement band of interest for a space-based observatory. The magnetic field can couple to the magnetic susceptibility and remanent magnetic moment from the test masses and disturb them from their geodesic movement. LISA Pathfinder carried on-board a dedicated magnetic measurement subsystem with noise levels of 10 $ \rm nT \ Hz^{-1/2}$ from 1 Hz down to 1 mHz. 
In this paper we report on the magnetic measurements throughout LISA Pathfinder operations. We characterise the magnetic environment within the spacecraft, study the time evolution of the magnetic field and its stability down to 20~$\mu$Hz, where we measure values around 200~$ \rm nT \ Hz^{-1/2}$ and identify two different frequency regimes, one related to the interplanetary magnetic field and the other to the magnetic field originating inside the spacecraft. Finally, we characterise the non-stationary component of the fluctuations of the magnetic field below the mHz and relate them to the dynamics of the solar wind. 
\end{abstract}




\input{include/introduction}

\input{include/instrument}
\input{include/magnetometer_DC}

\input{include/magnetometer_spectra}

\input{include/conclusions}

\input{include/acknowledgments}




\bibliographystyle{mnras}
\bibliography{library} 



\appendix

\input{include/appendix}

%


\bsp	
\label{lastpage}
\end{document}

%% file: include/author_list_extended_mnras.tex
\author[Armano et al.]{
M~Armano$^{a}$,
H~Audley$^{b}$,
J~Baird$^{c}$,
P~Binetruy$^{c}$\thanks{Deceased 30 March 2017},
M~Born$^{b}$,
D~Bortoluzzi$^{d}$,
E~Castelli$^{e}$, \newauthor
A~Cavalleri$^{f}$,
A~Cesarini$^{g}$,
A\,M~Cruise$^{h}$,
K~Danzmann$^{b}$,
M~de Deus Silva$^{i}$,
I~Diepholz$^{b}$, \newauthor
G~Dixon$^{h}$, 
R~Dolesi$^{e}$, 
L~Ferraioli$^{j}$,
V~Ferroni$^{e}$,
E\,D~Fitzsimons$^{k}$,
M~Freschi$^{i}$,
L~Gesa$^{l}$, \newauthor
F~Gibert$^{e}$, 
D~Giardini$^{j}$,
R~Giusteri$^{e}$,
C~Grimani$^{g}$,
J~Grzymisch$^{a}$,
I~Harrison$^{m}$,\newauthor
M-S~Hartig$^{b}$,
G~Heinzel$^{b}$,
M~Hewitson$^{b}$, 
D~Hollington$^{n}$,
D~Hoyland$^{h}$,
M~Hueller$^{e}$, \newauthor
H~Inchausp\'e$^{c}$ $^{o}$,
O~Jennrich$^{a}$,
P~Jetzer$^{p}$, 
N~Karnesis$^{c}$,
B~Kaune$^{b}$,
N~Korsakova$^{q}$, \newauthor
C\,J~Killow$^{q}$,
J\,A~Lobo$^{l}$\thanks{Deceased 30 September 2012},
L~Liu$^{e}$, 
J\,P~L\'opez-Zaragoza$^{l}$,
R~Maarschalkerweerd$^{m}$, \newauthor
D~Mance$^{j}$,
V~Mart\'{i}n$^{l}$,
L~Martin-Polo$^{i}$, 
J~Martino$^{c}$,
F~Martin-Porqueras$^{i}$,
I~Mateos$^{l}$, \newauthor
P\,W~McNamara$^{a}$,
J~Mendes$^{m}$, 
L~Mendes$^{i}$,
N~Meshksar$^{j}$,
M~Nofrarias$^{l}$,
S~Paczkowski$^{b}$, \newauthor
M~Perreur-Lloyd$^{q}$, 
A~Petiteau$^{c}$,
P~Pivato$^{e}$,  
E~Plagnol$^{c}$,
J~Ramos-Castro$^{r}$, 
J~Reiche$^{b}$, \newauthor
F~Rivas$^{l}$,
D\,I~Robertson$^{q}$,
D~Roma-Dollase$^{l}$,
G~Russano$^{e}$,  
J~Slutsky$^{t}$,
C\,F~Sopuerta$^{l}$, \newauthor
T~Sumner$^{n}$, 
D~Telloni$^{v}$,
D~Texier$^{i}$, 
J\,I~Thorpe$^{t}$,
C~Trenkel$^{u}$,
D~Vetrugno$^{e}$, 
S~Vitale$^{e}$, \newauthor
G~Wanner$^{b}$, 
H~Ward$^{q}$, 
P\,J~Wass$^{n}$,
D~Wealthy$^{u}$,
W\,J~Weber$^{e}$, 
L~Wissel$^{b}$,
A~Wittchen$^{b}$ \newauthor 
and P~Zweifel$^{j}$
\\
\\
\noindent \textit{Affiliations are listed at the end of the paper}
}

%% file: include/introduction.tex
\section{Introduction}
\label{sec:intro}

LISA Pathfinder~\citep{Anza05, Antonucci12} was an ESA mission with NASA contributions designed 
to test key technologies for the future gravitational-wave observatory in space, 
the Laser Interferometry Space Antenna (LISA)~\citep{Amaro17}. 
The main scientific goal for the mission was to reach a relative acceleration noise between two test masses in nominal geodesic motion of $3\times10^{-14}$ m s$^{-2}$ Hz$^{-1/2}$ at 1 mHz. 
The relevance of this measurement was not only its demand in terms of noise reduction but also the very low frequency measuring band, which introduces technological difficulties that can not be addressed by ground-based gravitational-wave detectors.

LISA Pathfinder was successfully launched on December 3rd, 2015, 
and started scientific operations after arriving at the Lagrange point L1 on March 1st, 2016. 
The mission plan included a six-month operations split between 
the two experiments on-board: the European LISA Technology Package (LTP) and the American Disturbance Reduction System (DRS). 
LISA Pathfinder achieved residual acceleration noise levels
of $\rm 1.74 \pm 0.01$ fm s$^{-2}$ Hz$^{-1/2}$ between 2\,mHz and 20\,mHz, 
and $\rm 60\pm10$ fm s$^{-2}$ Hz$^{-1/2}$ at 20\,$\rm \mu$Hz~\citep{Armano16, Armano18}, which went beyond the required perfomance and, hence, successfully demonstrated the technology to detect gravitational waves in space.

%
Apart from achieving a high level of free-fall, it is also important to understand the 
different contributions that will build up the instrument noise model. Therefore, one of the main objectives of LISA Pathfinder is to split up the noise into different contributions and help on the design of a suitable environment for 
future gravitational-wave detectors.
With that aim, LISA Pathfinder carried the so-called Data and
Diagnostics Subsystem (DDS) which includes 
a temperature measurement subsystem~\citep{Sanjuan07, Armano19_Temp}, 
a magnetic diagnostic subsystem~\citep{Diaz-Aguilo13} and 
a radiation monitor~\citep{Canizares09, Canizares11, armano2018GCR}. \\

In this work we focus on the results of the magnetic diagnostics and more specifically 
on the characterisation of the environment on-board the spacecraft during mission operations.
Understanding the variability of the magnetic environment is crucial for 
future space-borne gravitational-wave detectors since any magnetic perturbation has
a potential impact on the instrument performance through magnetic induced 
forces exerted on the test masses. With that aim, both interplanetary and 
platform originated magnetic fluctuations must be considered and characterised. 

This work is organised as follows. In section~\ref{sec.diagnostics} we introduce 
the nature of the magnetic forces that can perturb the test mass motion and 
describe the magnetic diagnostic system on-board, designed 
to study and disentangle this contribution. In section~\ref{sec.magnetometer}   
we describe the in-flight measurements and characterise the magnetic environment on-board 
as measured by the LISA Pathfinder magnetometers. Finally, we 
present our conclusions in section~\ref{sec.conclusions}.

%% file: include/instrument.tex
\section{The role of the magnetic environment in LISA Pathfinder}
\label{sec.diagnostics}

The geodesic motion of the test masses on-board LISA Pathfinder
is subjected to several sources of disturbance. Some of them 
can be corrected through the control loops that prevent, for instance, 
the push from solar radiation pressure to depart the mass from its 
free-fall orbit. Other sources of disturbance, however, can not be prevented and
need to be studied and characterised carefully since they will become
an inherent part of the instrument noise budget. This is 
the case of the forces of magnetic origin applied on the test mass. 

Indeed, an important contribution in the instrument 
budget is a force on the test mass that can arise
due to variations in the magnetic field on the test masses location,
originating a magnetic force on the test mass that could not be discriminated from
a force of gravitational origin. 

\subsection{Magnetic forces on the test mass}
\label{sec.magnetic_forces}
To understand the coupling of the test mass to magnetic induced forces one can consider the test mass inside the spacecraft as a magnetic dipole in a magnetic field. In such approximation, the test mass will sense a force given by
\begin{eqnarray}
\mathbf{F} = \left\langle \mathbf{M} \cdot \nabla \mathbf{B} \right\rangle V,
\label{eq.dipole}
\end{eqnarray}
where $\mathbf{M}$ is the intrinsic moment of the magnetic dipole, $\mathbf{B}$ the surrounding magnetic field and V the test mass volume. Here we use the brackets as an average over the test mass volume. 
For the sake of simplicity we do not take into account inhomogeneities or anisotropies in the composition of the test mass. A more detailed model of the effect of magnetic field on the test mass would consider also torques applied to the test mass~\citep{Diaz-Aguilo12}. Several considerations need to be taken into account for the
free-falling test masses inside the satellite. 
First, two contributions to the magnetic moment need to be distinguished: a first one from the intrinsic remanent magnetic moment ($\mathbf{M_r}$) and a second one coming from induced magnetic moment, 
which is proportional (through the magnetic susceptibility, $\chi$) to the applied magnetic field ($\mathbf{M_i} = \chi \mu_0^{-1}\,\mathbf{B}$).
Both contributions are determined by the composition of the test mass, which in the LISA Pathfinder case was a Pt (27\%) and Au (73\%) alloy. Eq. (\ref{eq.dipole}) therefore expands to 
\begin{eqnarray}
\mathbf{F} & = & \left\langle \left[ \mathbf{M_r} + \frac{\chi}{\mu_0}  \, \mathbf{B} \right]  \cdot \nabla \mathbf{B} \right\rangle V .
\end{eqnarray}

Since the main objective of the mission is to understand differential acceleration 
fluctuations down to the mHz, we further need to consider 
how fluctuations of the magnetic field will impact on the force. 
The different components to this contribution can be easily visualised if one 
splits both the magnetic field and the magnetic field gradient into a constant 
contribution and a term that varies with time, 
\begin{eqnarray*}
\mathbf{B} & = & \mathbf{B_{0}} + \delta \mathbf{B} \\
\nabla \mathbf{B} & = & {\nabla \mathbf{B}_{0}} + \delta {\nabla \mathbf{B}} ,
\end{eqnarray*}
which leads to several magnetically-induced force contributions 
\begin{eqnarray}
\mathbf{F} & = &  \left[ \left\langle   \mathbf{M_r} \cdot \nabla \mathbf{B_0} \right\rangle + \left\langle   \mathbf{M_r}   \cdot \delta \nabla \mathbf{B_0} \right\rangle \right] V \nonumber \\
 & + &   \left\langle \frac{\chi}{\mu_0} \left[   
    \mathbf{B_0} \cdot  \nabla \mathbf{B_0} + 
    \mathbf{B_0}  \cdot  \delta \nabla \mathbf{B} \right] \right\rangle V \label{eq.fluctuations_force_split}   \\ 
 & + &  \left\langle \frac{\chi}{\mu_0} \left[
    \delta \mathbf{B} \cdot \nabla \mathbf{B_0}  + 
    \delta \mathbf{B}  \cdot \delta \nabla \mathbf{B} \right] \right\rangle V \nonumber .
\end{eqnarray}
From Eq. (\ref{eq.fluctuations_force_split}) we clearly observe how the magnetic environment will induce a variety of effects on the test mass that include: constant force contributions ($\mathbf{B_0} \cdot \nabla \mathbf{B_0}$); terms that will couple the local field at the test mass position to the fluctuations of the gradient of the magnetic field ($ \mathbf{B_0} \cdot \delta \nabla \mathbf{B} $); couplings of the local magnetic field gradient to fluctuations in the magnetic field ($ \delta \mathbf{B} \cdot \nabla \mathbf{B_0}$) and coupling of fluctuations in the field and its gradient ($\delta \mathbf{B} \cdot \delta \nabla \mathbf{B} $). Hence, equally important for the experiment are both the monitoring of 
magnetic field and magnetic field fluctuations since
both can couple, through the test mass remanent magnetic moment and susceptibility, 
into spurious accelerations of the free-falling test masses. We will describe in the following the diagnostics subsystem used on-board to characterise these figures to later describe the results obtained with these sensors.

\subsection{The magnetic diagnostics subsystem}
\label{sec.instrument}
 
\begin{figure}
    \centering
       \includegraphics[width=0.5\textwidth]{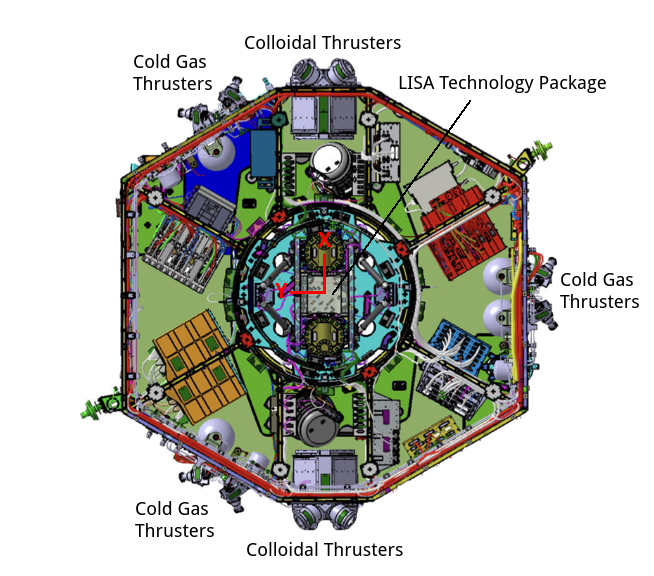} 
       \includegraphics[width=0.5\textwidth]{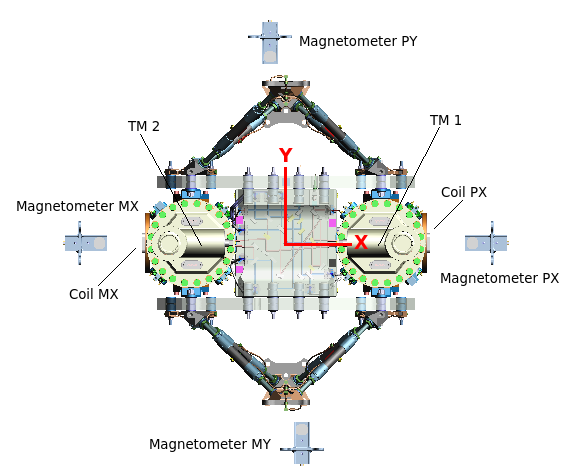} 
    \caption{\emph{Top:} X-Y plane view of the LISA Pathfinder spacecraft with the solar panel removed. The positions of the 3 sets of Cold Gas thrusters and the 2 sets of Colloidal Thrusters are indicated, as well as the LISA Technology Package in the center of the spacecraft. The LISA Pathfinder reference frame is shown in red, with its origin in the middle of the LISA Technology Package. 
    \emph{Bottom:} Zoom of the LISA Technology Package from the upper figure. Here we show the position of several items from the Data Diagnostics Subsystem, consisting in 2 coils and 4 tri-axial fluxgate magnetometers. In this notation `P' stands for `plus' and `M' for `minus', being each element named according to their position within the satellite reference frame. The position of both test masses is also indicated, as well as the LISA Pathfinder reference frame, like in the upper figure.}
\label{fig.LTPmagnetometers}
\end{figure}

The LISA Pathfinder magnetic diagnostics subsystem fulfils two related purposes. 
First, to measure the magnetic field and magnetic field gradient on-board and 
second, to create magnetic fields and gradients at the position of the test 
masses to study the contribution of the magnetic forces to the instrument noise budget. 
The subsystem is therefore composed by four tri-axial magnetometers and two induction coils 
to fulfil these two purposes, respectively.

The coils are capable of generating a controlled magnetic field at the location of the test masses. The two circular induction coils are located 85.5\,mm away from the test mass, each one attached to the external wall of each vacuum enclosure. The coils have a radius of 56.5\,mm and are built with 2400 windings of a copper alloy mounted on a titanium support with a dedicated high stability curent driver to ensure that high precision magnetic forces are produced~\citep{Diaz-Aguilo13}. The coils are aligned with the axis joining both test masses so that the generated magnetic field has axial symmetry. Experiments conducted to study magnetically-induced forces on the test masses during flight operations will be reported in a separate publication.
%

\begin{figure*}
\begin{center}
\includegraphics[width=0.9\textwidth]{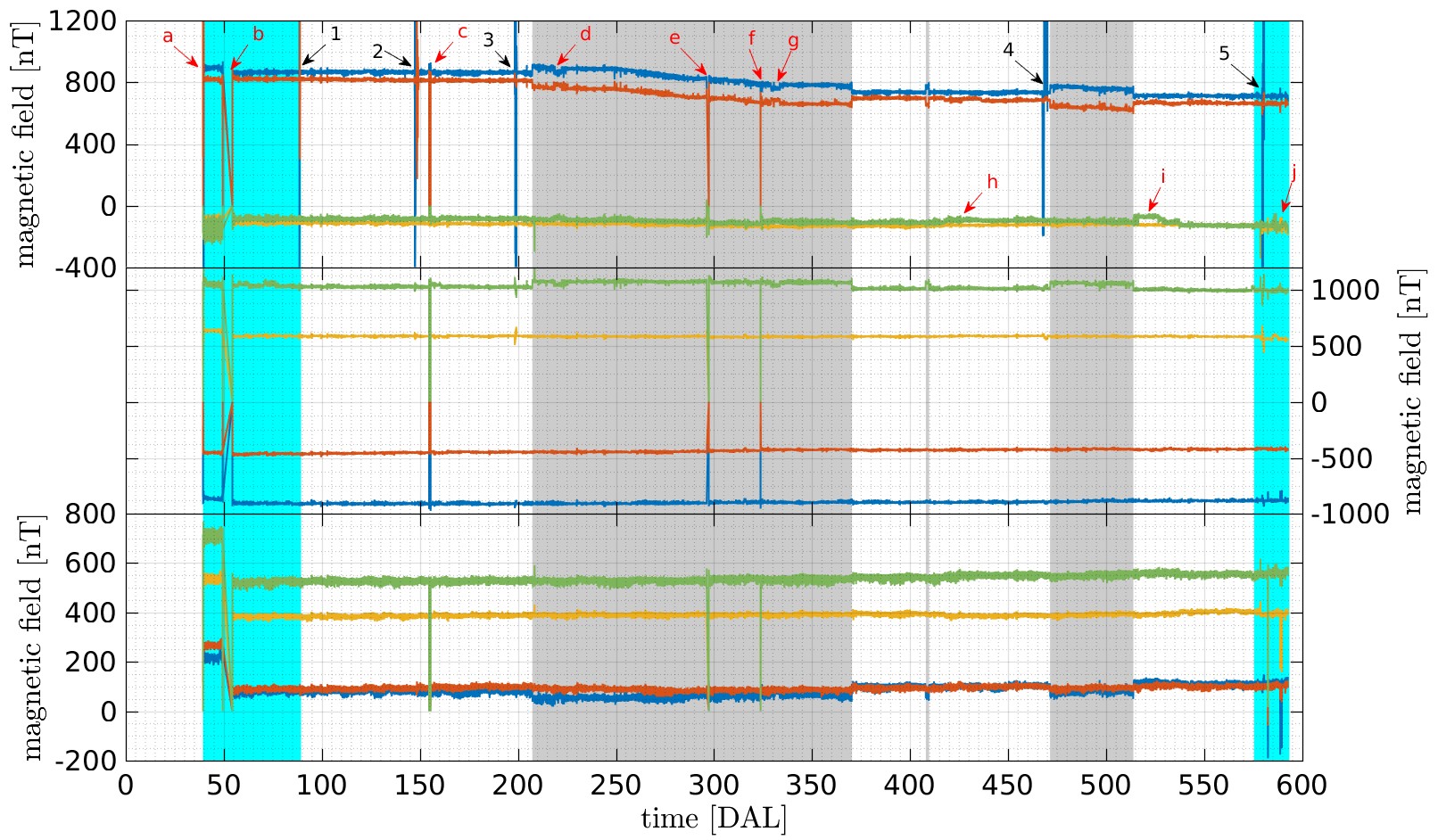}
\includegraphics[width=0.9\textwidth]{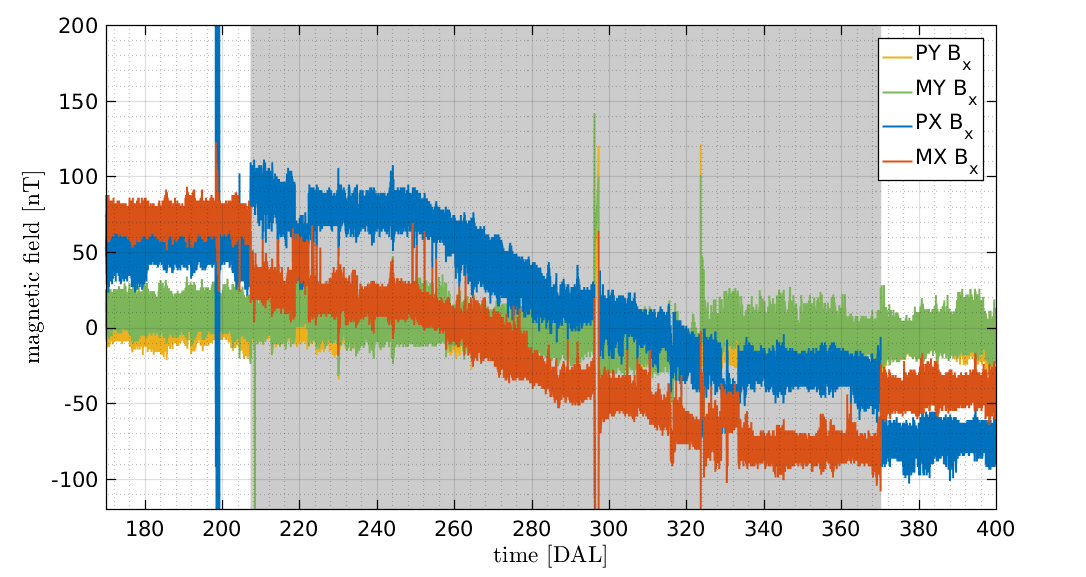}
\caption{\emph{Top:} Magnetic field measurements on-board LPF from launch date until the satellite passivation. The time axis is Days After Launch (DAL). The two cyan areas (DAL 40-90 and DAL 575-593) correspond to the commissioning and decommissioning periods, respectively. The 3 grey shadowed areas (DAL 210-370, DAL 414-415 and DAL 470-510) correspond to the DRS operations, and the rest correspond to LTP operations. 
From top to bottom pannel: the X, Y and Z component of the magnetic field, for magnetometers PX (blue), MX (red), PY(yellow) and MY (green).
In this notation, `M' stands for minus and `P' for plus. `X' and `Y' refer to axes on-board in which
the magnetometers are aligned, so PX and MX are the two closest magnetometers to the TMs. 
The arrows correspond to events related to important changes in magnetic field. These events are reported in  Tables~\ref{tbl.magnetic_injections} \&~\ref{tbl.magnetic_events}.
\emph{Bottom:} Zoom of X-component measurements for the four magnetometers during the first DRS operations period (Aug. - Nov. 2016). In this case, we removed the DC level from each channel to show all measurements in the same scale. See text for more detail.  \label{fig.timeseries}} 
\end{center}
\end{figure*}

The magnetometers are intended to sense with high precision the evolution of the magnetic field.
The magnetometers on-board LISA Pathfinder are fluxgate tri-axial magnetometers, namely TFM100G4-VQS from Billingsley with a sensitivity of $\rm 166.7 \mu V/nT$ and a typical root mean square noise figure of $\rm 20\,pT/Hz$ at 1 Hz.
In fluxgate magnetometers, each axis consists of a first sensing coil surrounding a second inner drive coil wound around a high permeability magnetic core material. Although providing low-noise measurements, because of the active magnetic core inside the sensor, these magnetometers have to be located far enough from the 
test mass position so they do not contribute as a source of magnetically-induced force on the test mass. 
All magnetometers channels are connected to a multiplexer after passing through an analog low-pass filter.  
The signal is then amplified by means of discrete operational amplifiers (OP471AY/QMLR) in an instrumentation amplifier configuration. Finally, it is digitally converted by a 16 bit analog to digital converter (7809LPRPFH). 
The sampling frequency is 50 Hz. Hence, since the input is multiplexed by 16 channels, the individual channel sampling frequency is 3.125 Hz.
In section \ref{sec.magNoise} we will come back to the two last components of the read-out chain, instrumentation amplifier and analog to digital converter, which determine the noise performance of the magnetometer read-out.  

In terms of data handling, the magnetometers data were downloaded with a dedicated telemetry package, which also included all the diagnostics on-board the satellite. 
This approach is a characteristic of LISA Pathfinder and differs from most of the scientific space missions where the diagnostics subsystem is considered only for house-keeping purposes. The rationale behind this is that, in gravitational-wave detectors, noise characterisation and noise hunting is crucial for the operation of the instrument. Hence, in our current case, environment sensors entail key information for the achievement of the scientific goal. 

%% file: include/magnetometer_DC.tex
\section{In-flight measurements}
\label{sec.magnetometer}

After a one-month cruise phase, LISA Pathfinder reached the L1 Lissajous orbit in January, 2016. The magnetometers on-board started collecting data on January 11th, at the start of the LTP commissioning phase and, apart from some short outages events, they acquired data uninterruptedly until the end of the scientific operations in July, 2017. 

Contrary to many other missions that carry their magnetometers at the end of a long boom to isolate the measurement from spacecraft interferences, magnetometers on-board LISA Pathfinder are located inside the spacecraft with the main objective of monitoring the magnetic field as close as possible to the test mass position.
At the same time, the magnetometers have to be sufficiently far away from the test masses since, as explained before, the fluxgate head contains a magnetically active component that could induce spurious forces on the test mass, therefore disturbing the main scientific measurement of the mission. 
The trade-off between these two conditions results in the locations shown in Fig.~\ref{fig.LTPmagnetometers}.

All magnetometers are located in the plane defined by the test masses and the optical bench. This configuration allows for an approximate estimate of the magnetic field gradient, a key factor in the magnetic noise contribution to the test masses free fall,  across two axes but it leaves the third one, Z in the spacecraft frame, unmeasured. It is worth stressing that the gradients estimated that way are unlikely to be representative of the magnetic field gradient at the test mass position, the reason being that local sources, as for instance the temperature sensors surrounding the test masses in the electrode housing, could be a potential source of local magnetic gradient ~\citep{Sanjuan09a}. For instance, for the worst case layout of these sensors, the magnetic field gradient across X between the 2 faces of the test masses could reach values around 10 $\mu$T m$^{-1}$, which would be orders of magnitude above the values we report in Appendix~\ref{sec.DCvalues} for the gradients computed across magnetometers. 
As we will show in the following, the DC value measured by each magnetometer is completely dominated by the spacecraft units while the fluctuations have both a spacecraft and an interplanetary contribution.

\subsection{Evolution of the measured on-board magnetic field}
\label{sec.DCmeasurements}

In the top panel of Figure \ref{fig.timeseries} we show the time series for the read-out of the four magnetometers on-board LISA Pathfinder since the magnetometers switch-on and until the end of the mission, in July 2017. We provide the magnetic field measurements in the X, Y and Z axes in the LTP reference frame, although the X-component contains most of the interesting features that we will discuss in the following. 

The absolute value of the magnetic field at each magnetometer location is dominated by the spacecraft contribution reaching values up to $1 \mu T$, far from typical interplanetary magnetic field values which would be of the order of 5-10\,nT. Most of the magnetic field measured by the magnetometers corresponds to the contribution of the cold gas micro-propulsion system, in particular to the magnets on the high-pressure latch valves. With an on-ground measured moment of ~950\,$\rm mA\,m^2$, they account for most of the magnetic field measured by the magnetometers~\citep{Armano15}. It is important to mention that since the cold gas thrusters are placed in different locations of the spacecraft the magnetic field they create do not necessarily add in the same direction.


The time series can be divided in six different segments which correspond to the different phases of the mission, namely comissioning, LTP operations, DRS operations, the associated mission extension for both experiments, and decomissioning. The most prominent features in the magnetic field timeline are experiments with the coils to characterise the magnetic contribution to the test mass free-fall. These experiments are listed in Table \ref{tbl.magnetic_injections}. Other features that affected the magnetic environment and can be identified in Figure \ref{fig.timeseries} are listed in Table \ref{tbl.magnetic_events}. In the latter case, these are not associated to the activation of the coils on-board (except for event `a', which was a check of the correct functioning of coil 2) but to the operation of the spacecraft itself, i.e. from changes in configuration to identified anomalies during operations. For the sake of completeness, the mean values of the magnetic field DC for different configurations of the spacecraft can be found in Appendix~\ref{sec.DCvalues}.

In terms of the magnetic environment, we notice that DRS operations had an impact in the spacecraft magnetic environment. As seen in the bottom panel of Figure~\ref{fig.timeseries}, the mean value of the X component shows a steady decrease of ~150\,nT starting around Aug. 8th and ending around Nov. 13th. for the two magnetometers located near the test masses, i.e. magnetometers PX and MX in the bottom panel of Figure~\ref{fig.LTPmagnetometers}. This period is coincident with the start of operations of the DRS instrument. Apart from this long drift, in magnetometers PX and MX we also observe a DC increase of about 40\,nT when we turn on the DRS system and the same decrease when we turn it off. This DC change should be related to some units being switched on and off when we change the control system of the satellite.

The reason of this long term trend observed in the magnetic field on-board can be explained by the differences between LTP and DRS micro-propulsion systems.
The LTP micro-propulsion system consists of three clusters each featuring four cold gas thrusters~\citep{Armano19}.
The thruster system uses high-pressure Nitrogen propellant to provide very small impulses with a thrust range of 1--500\,$\rm \mu N$. Four Nitrogen tanks store the gas at 310 bar with a maximum capacity of 9.6 kilograms of Nitrogen.
The DRS micropropulsion system is composed by two Colloid Micro-Newton Thruster (CMNT) clusters~\citep{Ziemer06}. Each cluster includes four complete and independent thruster units. For the CMNT subsystem, thrust is adjustable from 5--30\,$\rm \mu N$ by changing the beam voltage  and/or propellant flow rate that determines the beam current. In this case, the propellant is stored in four electrically isolated stainless steel bellows compressed by four constant force springs set to supply four micro-valves with propellant at approximately 1\,atm of pressure. In terms of the spacecraft magnetic field, this sets a relevant difference between both thrust subsystems since the continuous operation of the CMNT subsystem leads to a steady displacement of the bellows inside the storage tank.
Although not having a direct impact on the mission operations in terms of gravitational pull, it is precisely this displacement of the stainless steel bellows occurring during the depletion of the fuel tanks the one that originates the observed magnetic field change. Indeed, given the known geometry and measured magnetic properties of the CMNT we can estimate the impact of the operation of these thrusters on the magnetic environment. The CMNT are aligned in the direction joining both test masses (X direction) as can be seen in the top panel of Fig. \ref{fig.LTPmagnetometers}. They are located at a distance of 28\,cm with respect to the two magnetometers that are located along this same axis, the PX and MX magnetometer in the notation of the bottom panel of Fig. \ref{fig.LTPmagnetometers}. The measurement on-ground reported a magnetic moment for these units of 209\,$\rm mA\,m^2$ (in modulus) which would produce a magnetic field of 1.1\,$\rm \mu T$ on the position of the closest magnetometer to each CMNT according to the equation of the magnetic field produced by a magnetic dipole:
\begin{eqnarray}\label{eq.magfield_dipole}
    \mathbf{B(r)}=\frac{\mu_0}{4\pi}\left( \frac{3\mathbf{r}(\mathbf{m}\cdot\mathbf{r})}{\mathbf{r}^5} - \frac{\mathbf{m}}{\mathbf{r}^3} \right)
\end{eqnarray}
with $\mathbf{m}$ being the magnetic moment of the source and $\mathbf{r}$ a vector going from the center of the magnetic dipole to the position where the magnetic field is being measured. If we now assume that the overall variation in the magnetic field measured in the X direction during DRS operations is due to the displacement of the bellow inside the CMNT valves, this would imply a 1\,cm displacement which is compatible with the geometry of the valve and the amount of propellant being expelled during this period.

\begin{table}
\begin{tabular}{lr}
\hline
Event& Date (DAL) \\  
\hline
(1) Comissioning injec. in TMs 1 \& 2 & 28 Feb'16 (87) \\
(2) Mag. exps. in TMs 1 \& 2  & 27 April'16 (146) \\
(3) Mag. exps. in TM 1        & 18 Jun'16 (198) \\
(4) Mag. exps. in TM 1        & 14 Mar'17 (467) \\
(5) Decomissioning injections & 04 Jul'17 (579) \\
\hline
\end{tabular}
\caption{Dates associated with magnetic experiments on-board LISA Pathfinder. In parenthesis we include the Days After Launch (DAL).}
\label{tbl.magnetic_injections}
\end{table}

\begin{table}
\begin{tabular}{lr}
\hline
Event& Date (DAL) \\  
\hline
(a) Coils check & 11 Jan'16 (39)\\
(b) Propulsion module released & 22 Jan'16 (50)\\
(c) DMU SW crash & 5 May'16 (154)\\
(d) Cluster-2 DCIU anomaly & 9 Jul'16 (219)\\
(e) LTP safe mode & 24 Sep'16 (296)\\
(f) DMU SW crash and reboot & 21 Oct'16 (296)\\
(g) Thruster-4 anomaly & 27 Oct'16 (329)\\
(h) Cooling down & 23 Jan'17 (417)\\
(i) Cooling down & 29 Apr'17 (513)\\
(j) High pressure latch valves switch & 13 Jul'17 (588) \\
\hline
\end{tabular}
\caption{Dates associated with events that impacted the magnetic environment on-board LISA Pathfinder. In parenthesis we include the Days After Launch (DAL).}
\label{tbl.magnetic_events}
\end{table}

%% file: include/magnetometer_spectra.tex
\subsection{Fluctuations of the on-board magnetic field}
\label{sec.spectra}

Given that any varying magnetic field or magnetic field gradient will couple into the motion of the test masses, 
understanding the origin of the fluctuations of the magnetic field measured on-board is an important output of LISA Pathfinder for future gravitational-wave detectors in space. In the previous section we have seen how the electronic units on-board the satellite are the dominant contribution to the magnetometers DC measurement since the interplanetary contribution (typically of the order of 5-10\,nT) is at least one order of magnitude below the values reported, for instance, in Table \ref{tbl.DCvalues}. The situation is different when we study the variations of the magnetic field. A wide variety of phenomena can 
produce a varying magnetic field and, as we show in the following, both fluctuations originated by the spacecraft and by the interplanetary magnetic field are relevant to understand the magnetic field spectra measured by LISA Pathfinder. 

We divide the three sections below as follows. First, we provide a characterisation of the different magnitudes describing the fluctuations of the magnetic field on-board to then focus on the physical mechanisms describing these fluctuations for two different frequency regimes that we distinguish in our data.

\begin{figure*}
\includegraphics[width=0.49\textwidth]{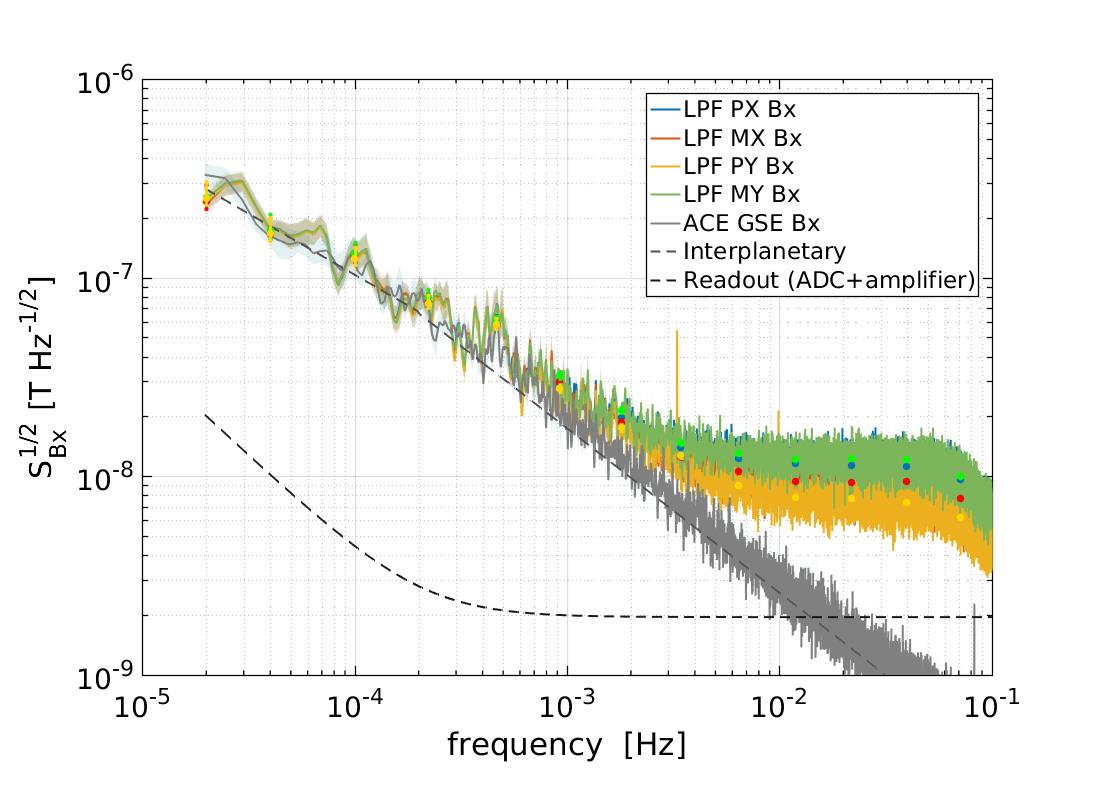}
\includegraphics[width=0.49\textwidth]{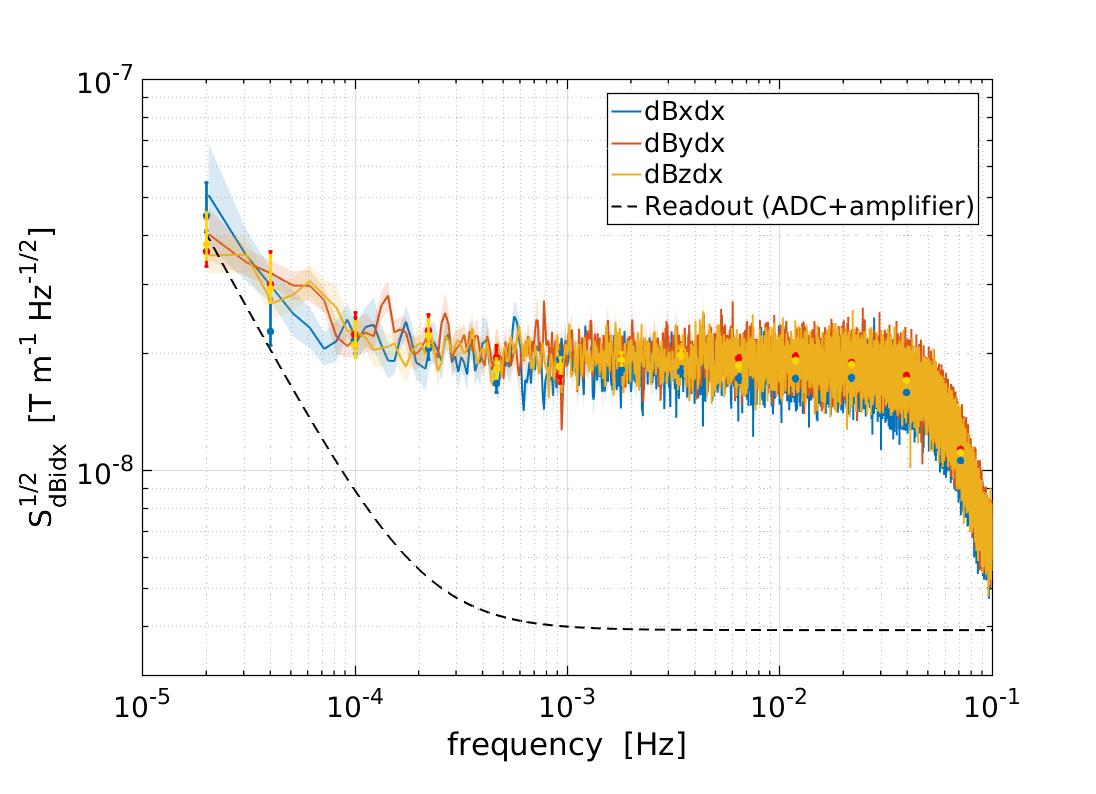}
\includegraphics[width=0.49\textwidth]{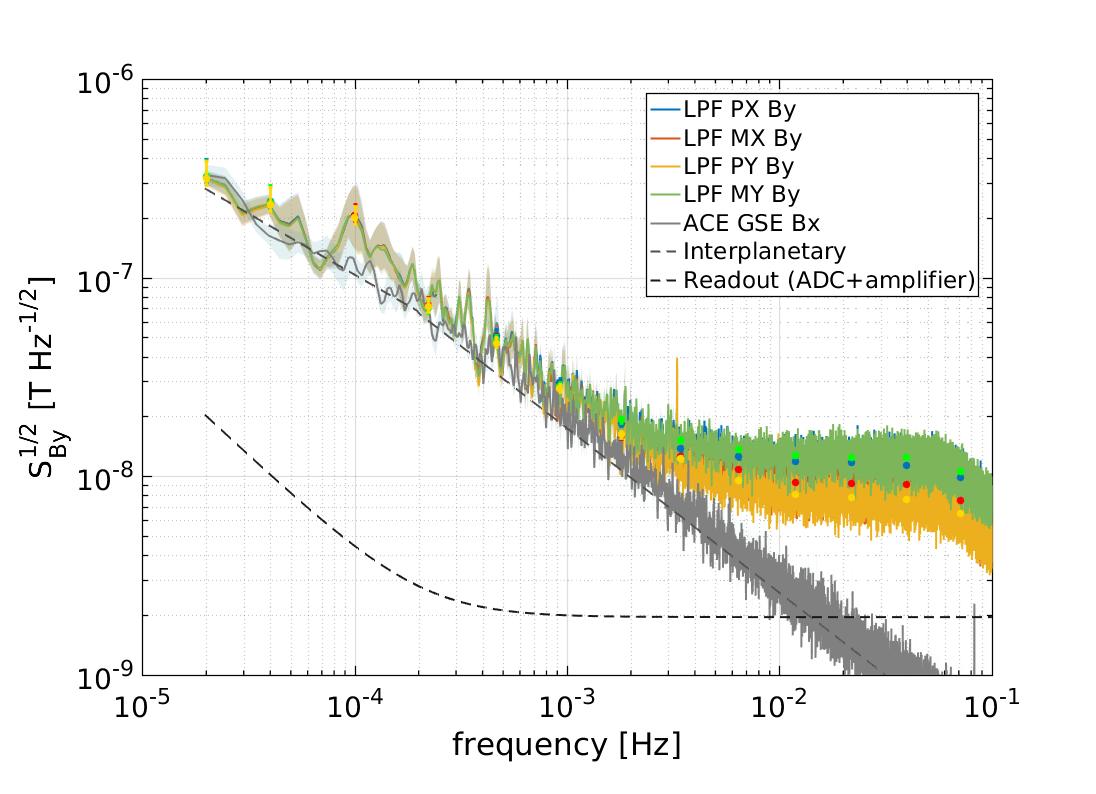}
\includegraphics[width=0.49\textwidth]{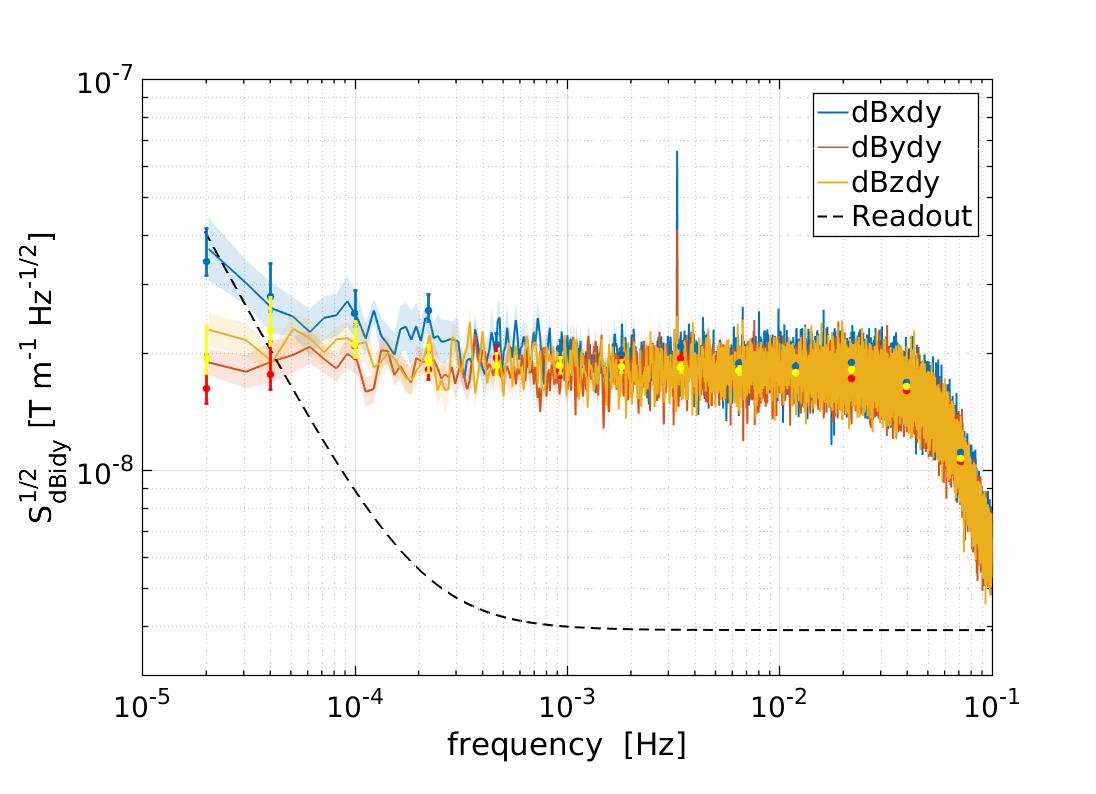}
\includegraphics[width=0.49\textwidth]{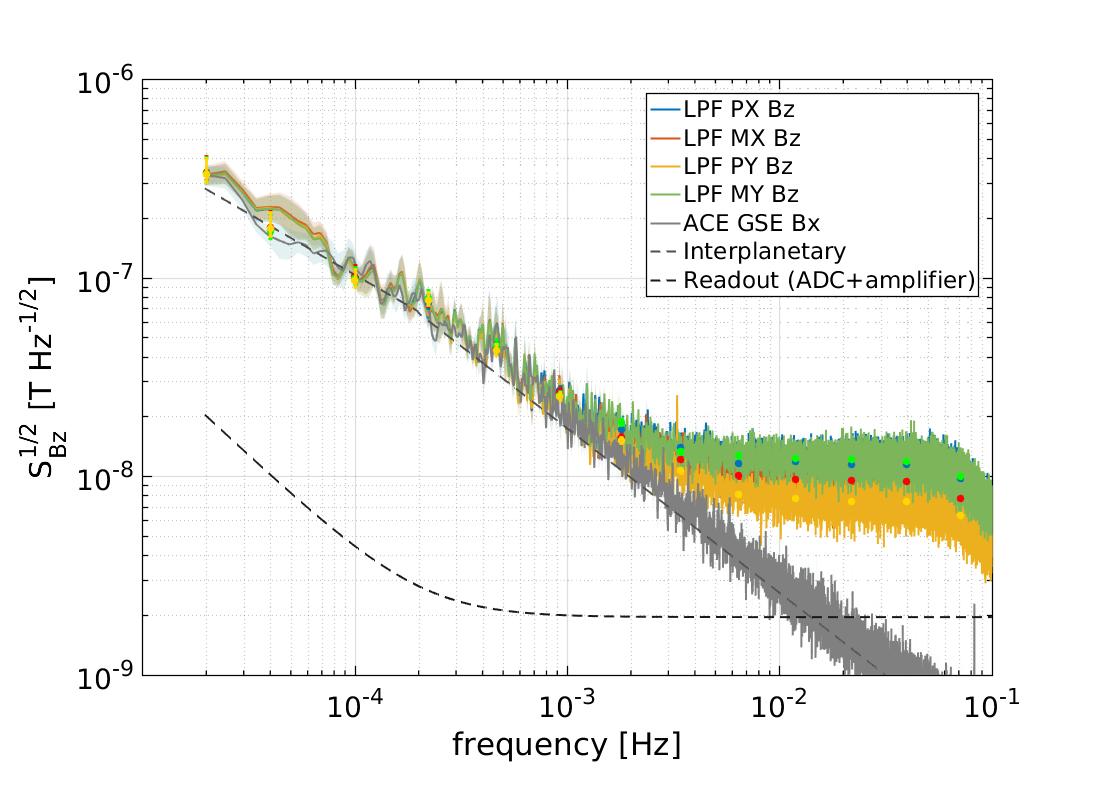}
\includegraphics[width=0.49\textwidth]{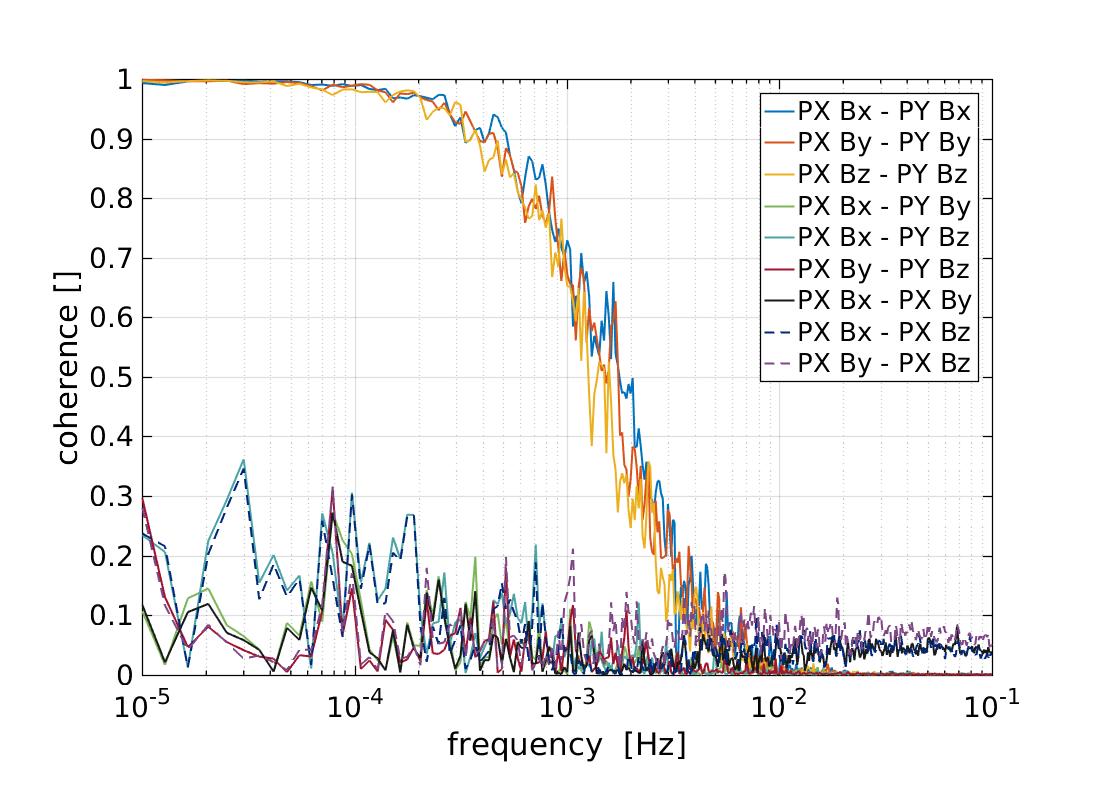}
\hspace{8.5cm}
\caption{Characterisation of the magnetic field environment on-board for the period from 13 Feb'17 to 2 Mar'17.
 \emph{Left:} Spectrum of the, from top to bottom, X, Y and Z components of the magnetic field from the four LPF magnetometers and from the ACE magnetometer. Dashed lines correspond to the contribution to LPF magnetometers coming from interplanetary magnetic field and electronics. 
 \emph{Right:} Spectrum of the gradient of the magnetic field along the X-axis (uppermost panel) and Y-axis (middle panel) on-board LPF. The dashed lines correspond to the contribution coming from electronics to LPF magnetometers. The bottom panel corresponds to the coherence function between the three axes magnetic field as measured in independent magnetometers on-board the satellite.}
\label{fig.PSD}
\end{figure*}

\subsubsection{Characterisation of magnetic field fluctuations on-board}
\label{sec.MagFieldcharact}

The fact that the four magnetometers of the magnetic diagnostics subsystem enclose 
the main instrument enables the direct comparison 
between different read-outs to disentangle spacecraft from interplanetary contributions. 
To do so, we use the coherence function which statistically quantifies 
common fluctuations between two time series.
In the bottom right panel of Figure~\ref{fig.PSD} we show the coherence function between the different channels of the magnetometers pair PX and PY, although the results shown are equivalent for each couple of magnetometers analysed. 
As shown in the figure, coherence between orthogonal axes is suppressed in the whole band 
while correlation between measurements on the same axes show a steep increase below 5\,mHz.
Given the low frequency of the coherent magnetic field fluctuations and the location of the magnetometers, these results points towards the interplanetary magnetic field fluctuations as the leading contribution in the sub-mHz frequency regime. On the other hand, fluctuations above the mHz region would be dominated by read-out electronics. This is further confirmed by the following analysis.  


To evaluate the magnetic field and magnetic field gradient fluctuations we compute the amplitude spectral density (the square root of the power spectral density) by means of the Welch averaged periodogram~\citep{Welch67}. 
We use data segments of 56\,h and apply a Blackman-Harris window to prevent spectral leakage.
To make sure that the window is not biasing our estimate we remove the lowest four frequency bins of the spectra.
To compute the dots in the spectra we performed an averaging in the frequency domain, see the method described in the supplemental material of \cite{Armano18} for a further explanation.
All the analysis and data post-processing were performed with LTPDA~\citep{Hewitson09}, a MATLAB toolbox designed for the analysis of the LISA Pathfinder data. 
 
The spectra of magnetic field fluctuations are shown in the left panel of Figure~\ref{fig.PSD}, from top to bottom, for the X, Y and Z components of the magnetic field, respectively. In this case, we notice that the two frequency regimes previously observed in the coherence function also appear with differentiated spectral behaviour above and below 5\,mHz. Indeed, while the sub-mHz fluctuations show an almost constant spectral index, fluctuations above the mHz are flat and uncorrelated. Moreover, the magnitude of sub-mHz fluctuations is equal for all four magnetometers whereas in the high frequency regime different magnetometers show a different noise plateau.
In the different panels of Figure~\ref{fig.PSD} we show some fits to the data (the small dots). We will describe them in detail in section~\ref{sec.nonstationarities}, where we will show how changes in some parameters of the interplanetary media affect the spectrum of the magnetic field.




As we described in Sec.~\ref{sec.instrument}, not only the gradient of the magnetic field but also the fluctuations of this gradient can contribute as a force exerted on the test mass. Thus, we took advantage of the configuration of the magnetometers on-board to estimate the magnetic field gradient from the difference between magnetic field measurements on opposing sides of the LTP. Opposing magnetometers are separated by 0.65 cm in the PX-MX case, and 0.54 cm in the PY-MY case. Check bottom panel of Figure \ref{fig.LTPmagnetometers} for a clear view of the setup. 
All magnetometers in the spacecraft are placed in the same X-Y plane and, therefore, no estimates of the Z gradients were possible. In the right top and right middle panels of Figure \ref{fig.PSD} we show the amplitude spectral density of the fluctuations of the magnetic field gradient across the X and Y axis, respectively, of LISA Pathfinder for the three components of the magnetic field. 
The spectra of the magnetic field gradient are flat at $\rm 20\,nT \ m^{-1} Hz^{-1/2}$ down to about 0.1 mHz, to even smaller frequencies than in the case of the magnetic field fluctuations. In this case, the noise measured seems to be in accordance with our electronics noise (see Section~\ref{sec.magNoise} for more details), and thus, we are limited by our instrument to measure any fluctuation smaller than this value. It should be reminded than this value is not likely to be representative of the magnetic field gradient at the test mass position, since any local source of magnetic field close to the test mass could be a potential source of magnetic field gradient that would not be measured by our pair of magnetometers if they are too far away from the mentioned source.

Although we have previously described the smooth shape of the magnetic field spectra, there are also some spectra lines appearing in the panels in Figure \ref{fig.PSD}. In some of them a line at 3.3 mHz appears in the PY magnetometer channels, with its corresponding harmonic at 10 mHz. Moreover, the signal appears to be stronger in the X component of the magnetic field.
We can not confirm the physical origin of these lines. However, since it is clearly visible in the magnetic field gradient across the Y-axis, this points out to a local origin and excludes any source coming from the interplanetary media, which would be sensed equally by all our magnetometers. The distribution of the units and the magnetometers on-board points towards the On-Board Computer (OBC) as the probable source of this magnetic field tone.

\subsubsection{Fluctuations in the sub-millihertz: interplanetary magnetic field contribution}
\label{sec.lowFreqFluct}
The interplanetary magnetic field measured by our magnetometers is imprinted on the solar wind plasma that surrounds the spacecraft and travels through the interplanetary media.
Plasma fluctuations in interplanetary space have been successfully described in the framework of the classical Kolmogorov turbulence~\citep{Kolmogorov1941}. In this framework, energy injected into the interplanetary plasma at large scales is transferred by non-linear interactions to microscales where it is finally dissipated, thus heating the plasma. The low-frequency part of the magnetic field and plasma-velocity power spectra often exhibits a clear $f^{-1}$ scaling, from DC up to frequencies of about $10^{-4}$ Hz in the fast Alfv\'enic wind~\citep{Bruno2009}, where the turbulent energy cascade becomes active. It is worth noting that, contrary to the fast solar wind, in the low-speed streams the injection range may cover a smaller range of frequencies and sometimes not to be present at all~\citep{Bruno2019}. At frequencies higher than $10^{-4}$ Hz but below the ionic break that occurs around 0.1--1 Hz, we find what we call the inertial range. In this range, the solar wind is in a state of fully developed turbulence, where the magnetic energy spectrum has a well defined Kolmogorov $f^{-5/3}$ spectrum~\citep{Bruno2013}. At frequencies higher than the ionic break the Kolmogorov spectrum breaks down and the magnetic fluctuations display a steeper $f^{-7/3}$ power-law spectrum, up to frequencies of about 100 Hz~\citep{Sahraoui2009}, where dissipation processes at proton scales take place. At even higher frequencies sometimes the spectrum steepens even more, roughly described by a further power law~\citep{Sahraoui2009} with a scaling exponent between $f^{-3.5}$ and $f^{-5.5}$, or perhaps by an exponential decay~\citep{Alexandrova2012}. Since these scales suffer for a lack of spacecraft measurements, a clear indication cannot be provided. This region of frequencies has been indicated as a range where collisionless dissipative mechanisms are efficiently at work at electron scales~\citep{Sahraoui2009} \&~\citep{Alexandrova2012} \&~\citep{Goldstein2015}.

In order to check our measurements with previous characterisation of the solar wind, in the left panels of Figure \ref{fig.PSD} we compare the amplitude spectral density of magnetic field fluctuations obtained during a LISA Pathfinder noise run to a set of data obtained from the Advanced Compton Explorer (ACE)~\citep{Smith98} in the same period of time. The ACE mission monitors different parameters of the solar wind by means of a suite of instruments on-board and, as LISA Pathfinder, follows a Lissajous orbit around L1. It is therefore a useful dataset with which to compare our measurements. ACE data is shown in Geocentric Solar Ecliptic Coordinates (GSE) system, which has its X-axis pointing from Earth towards the Sun and its Y-axis is chosen to be in the ecliptic plane pointing towards dusk. It is worth mentioning that even though both satellites are orbiting around L1, the distance separating them can be of the order of $10^5$--$10^6$ kilometers. However, we can safely compare the fluctuations of the magnetic field between both satellites at frequencies around the 20--50 $\mu$Hz, which is the band in which we will focus our analysis. As we will discuss in more detail, the typical velocity of around 200--500 km s$^{-1}$ of the solar wind guarantees that fluctuations in this frequency range have a coherence length greater than the distance between both spacecraft. 


Our results show that, for frequencies below 3\,mHz, the amplitude of fluctuations measured in LISA Pathfinder are in agreement with those measured by ACE during the same period of time. The spectral index obtained by both instruments is in agreement with previous characterisations of the spectra of interplanetary magnetic field fluctuations corresponding to the inertial range, as we will see in detail in section~\ref{sec.nonstationarities}. From our analysis we confirm that whilst the absolute value of the magnetic field on-board is dominated by the units inside the spacecraft, the fluctuations of the magnetic field are instead dominated by the fluctuations of the interplanetary magnetic field. 
At the same time, this corroborates the results obtained in the bottom right panel of Figure \ref{fig.PSD}. Since all magnetometers are measuring the interplanetary contribution, fluctuations of the magnetic field are completely correlated in the low frequency range for those channels measuring the magnetic field in a given direction. This correlation decays if we compare measurements of the magnetic field in transverse directions. As expected, the correlation between fluctuations of the magnetic field disappears for frequencies above 3\,mHz. As we show in the following, this frequency range is dominated by electronic noise, i.e. with no common correlation between channels.

\subsubsection{Fluctuations above the millihertz: fluxgate read-out electronics }
\label{sec.magNoise}

In section \ref{sec.instrument} we provided a description of the read-out chain of the magnetometers. The noise  analysis for this chain shows that the amplifier and the analog-to-digital converter which will limit the performance of our sensor at low and high frequencies, respectively. If we take into account the different components in the read-out chain and the noise figures in the data sheet values, we obtain the contribution shown in the different panels of Figure~\ref{fig.PSD}. On the one hand, in the low-frequency region, the noise follows a $f^{-1}$ spectrum produced by the instrumentation amplifier. On the other hand, in the high-frequecy region, the spectrum is flat due to the analog-to-digital converter which sets a limit of $S^{1/2}_{ADC} \simeq 2$~nT~Hz$^{-1/2}$ at the high frequency band.
In the left panels of Figure \ref{fig.PSD} we show, for the three different axes, the noise floor measured by the four magnetometers on-board during the period. On-board magnetometers measured a noise level above 3\,mHz that differs for each magnetometer in a range that goes from $\rm 7\,nT \ Hz^{-1/2}$ in the PY magnetometer to $\rm 11\,nT \ Hz^{-1/2}$ in the MY magnetometer, all of them above the expected $\rm 2\,nT \ Hz^{-1/2}$.
Considering that the amplitude of the interplanetary fluctuations decay as $f^{-1.65}$ and taking into account that the observed noise is not correlated between the four magnetometers --see bottom right panel of Figure \ref{fig.PSD}, we conclude that the read-out electronics must be the source of this excess noise. We have investigated this by focusing on the electronics design of the magnetometer read-out. 
Our analysis shows that this noise contribution could be assigned to a common-mode noise at the input of the instrumentation amplifier which can be originated due to the lack of common ground between magnetometer and electronics read-out.
We have experimentally tested this hypothesis by means of an engineering model of the LISA Pathfinder Data Management Unit (DMU) \citep{Canizares11} which included the Data Acquisition Unit (DAU) unit together with a flight model fluxgate magnetometer.
With this setup we have verified that the measured noise plateau can vary from the design $\rm 2\,nT \ Hz^{-1/2}$ if both units are not commonly grounded.
Hence, we conclude that the observed excess noise above 3\,mHz could be assigned to this issue. 
Although it is not possible to assess the exact contribution to the noise budget due to this effect, a worst case estimate sets a value of 90\,$\rm mV \ Hz^{-1/2}$ for the required fluctuations at the input of the instrumentation amplifier in order to explain the excess observed by our magnetometers. This value is relatively high compared to our read-out voltage noise. Since it is not possible to measure the common-mode noise at the spacecraft, we are only establishing an upper bound without discarding other possible contributions to the excess noise in the high-frequency band. It is important to stress here that the main objective of the magnetometers on-board LISA Pathfinder was to track slowly varying magnetic fields that are the ones that can have an impact in the dynamics of the free falling test mass. Hence, an excess noise in the high frequency range, though unexpected, does not have an impact on the scientific objectives of the magnetic diagnostic subsystem, but if needed, it could be corrected for a future space-based gravitational wave mission. 
 
\begin{figure*}
\begin{center}
\includegraphics[width=0.49\textwidth]{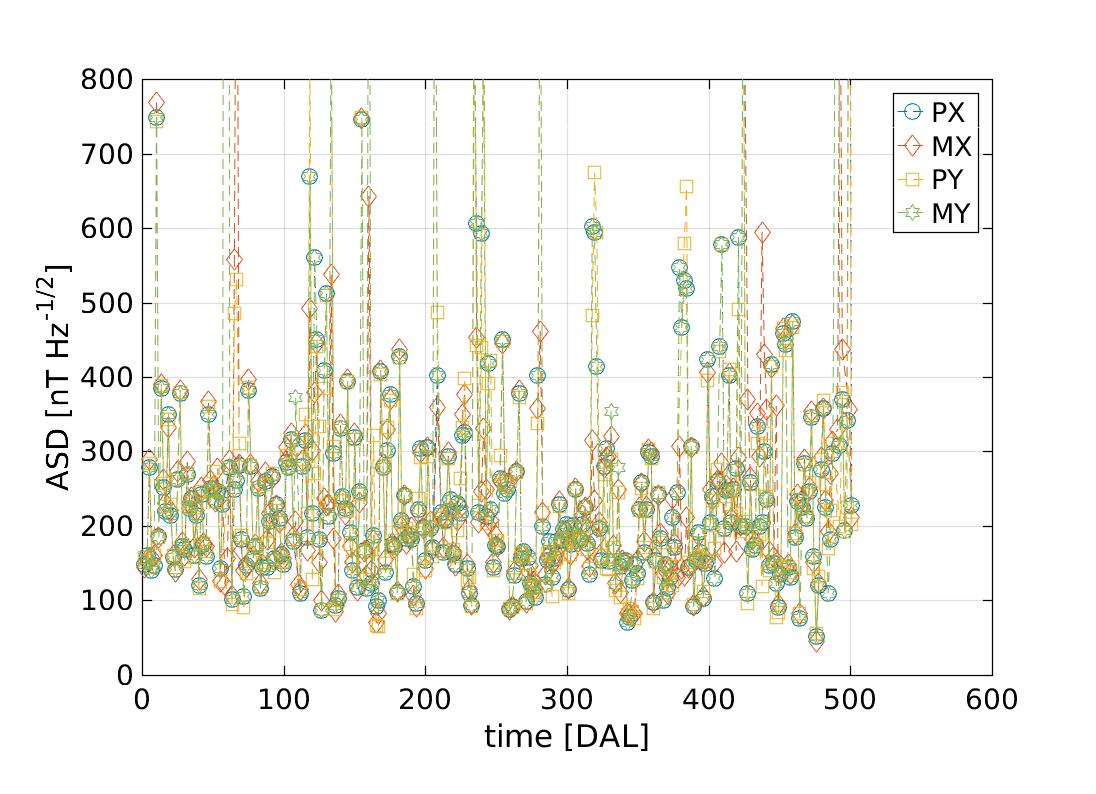} 
\includegraphics[width=0.49\textwidth]{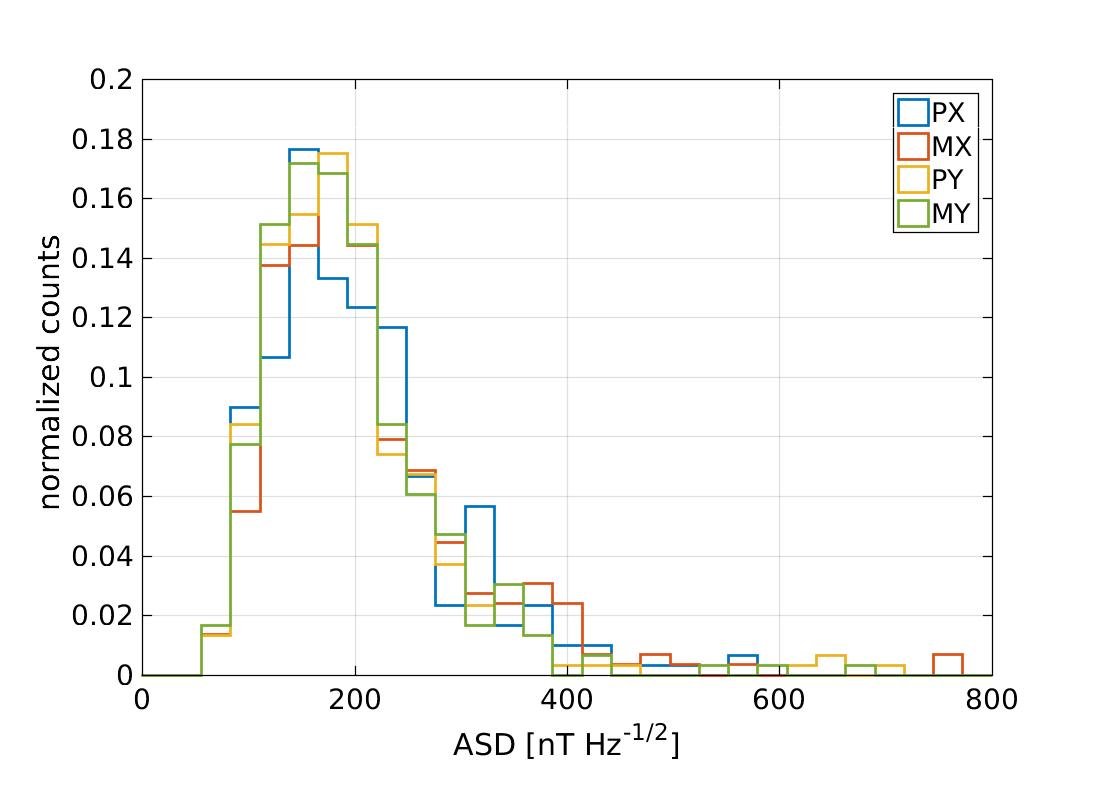} 
\includegraphics[width=0.49\textwidth]{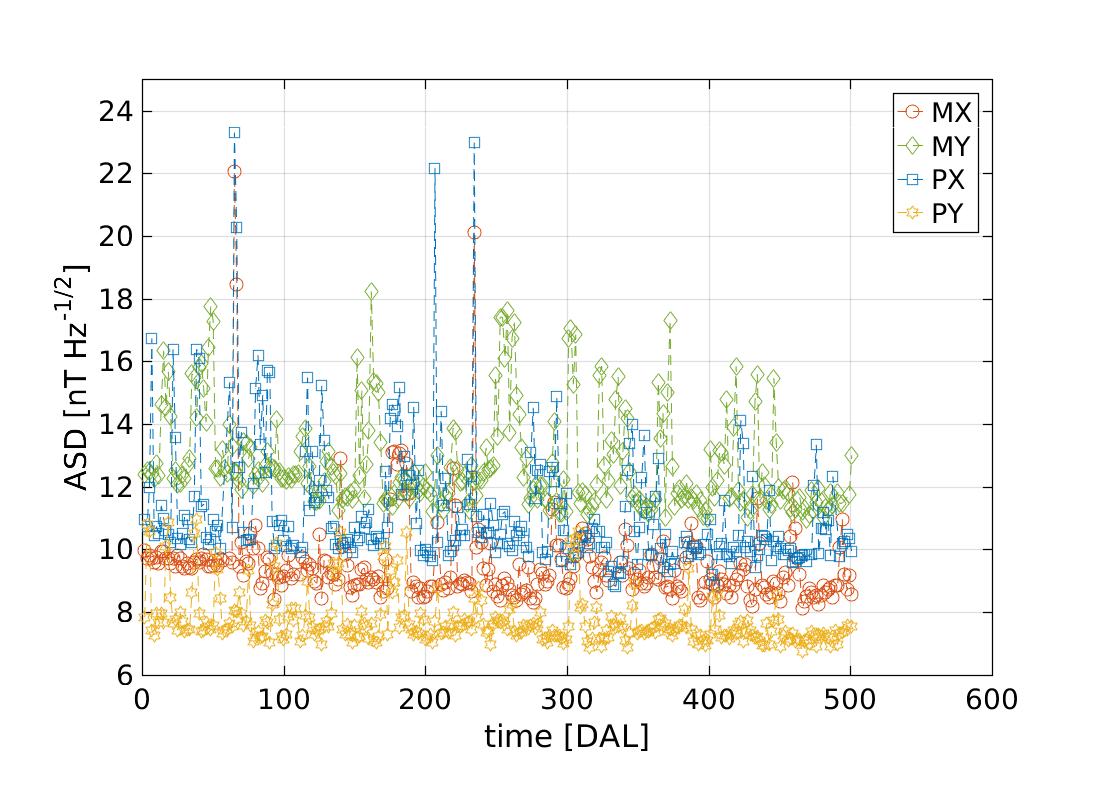} 
\includegraphics[width=0.49\textwidth]{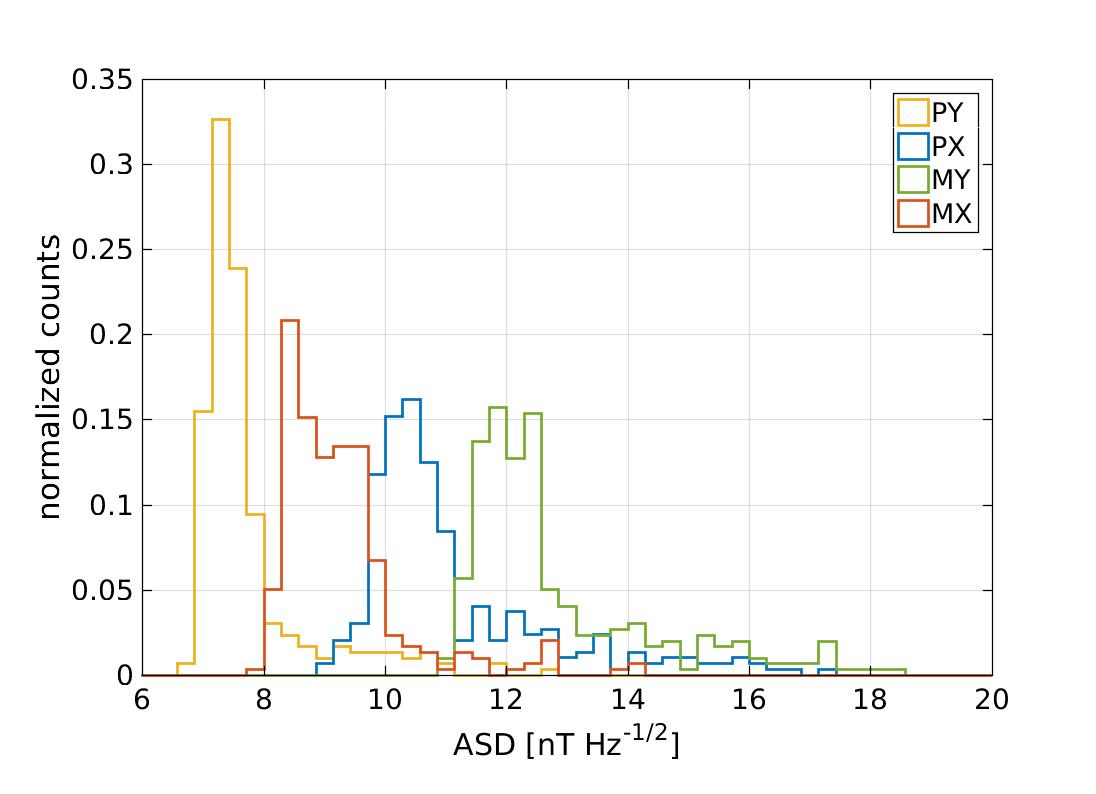} 
\caption{Time evolution and statistical distribution of the X-component of the magnetic field fluctuations as measured on-board, for two different frequency regimes. Colours correspond to magnetometers: PX (blue), MX (red), PY (yellow) and MY (green). Time is indicated in days after launch (DAL), while ASD stands for Amplitude Spectral Density (ASD). \emph{Top:} Fluctuations in the frequency range  $\rm 20 < \ \it f < \rm 50 \ \mu Hz$. \emph{Bottom:} Fluctuations in the frequency range   $\rm20 < \ \it f < \rm 40 \, mHz$.
 \label{fig.power}}
\end{center}
\end{figure*}

\subsection{Non-stationarities in the magnetic field fluctuations}
\label{sec.nonstationarities}

Until now we have based our analysis on the amplitude spectral density, 
which entails the information of the 
magnetic field in the frequency span of interest.
The amplitude spectral density effectively describes the fluctuations during 
a fixed period of time. Hence, it would only be a complete statistical description if the environment on-board LISA Pathfinder were stationary. 
This is obviously not the case. On the contrary, 
several situations can induce abrupt changes in the measured magnetic field.
In Table~\ref{tbl.magnetic_events} we provided a series of events that we identified in the 
magnetic field timeline. Since these were associated to satellite operations these events 
are both easily identified and, eventually, removed from the dataset through post-processing.  
A second, more complex, class of non-stationarity is the one associated to 
the interplanetary magnetic field. In section \ref{sec.lowFreqFluct} we described 
the origin of the spectral index of the observed magnetic field fluctuations 
in the low frequency range and its relation with the solar wind speed. 
A rich variety of interplanetary structures is superposed to the unperturbed solar wind plasma~\citep{Kilpua2017} \& ~\citep{Richardson2018}. Structures such as corotating interaction regions, interplanetary shocks, magnetic clouds or heliospheric current sheet crossings will induce variations in the measured magnetic field on-board the spacecraft. 



Since our DC magnetic field is completely dominated by the spacecraft components: around $\rm \sim 1 \mu T$, compared with the $ \rm \sim 1nT$ coming from the interplanetary media, we can not measure the absolute value of the interplanetary magnetic field in these cases. Nevertheless, if the variations are strong enough (of the order of $\rm \sim 10 nT$, for example) and especially if we see them in all four magnetometers, we could indeed assign an interplanetary origin to them.
Accurate detection of the magnetic imprint of these events requires the magnetometers to be isolated from spacecraft
contributions. This is the case for dedicated space weather mission which place the magnetometers at the end of a deployable boom, such as in the case of ACE~\citep{Smith98} or WIND~\citep{Lepping93} missions. 
Moreover, these missions contain a suite of instruments that allow a complete characterisation of the plasma, tracking parameters such as the solar wind speed or the number density of the plasma that we will refer in the following. 
Nonetheless, it is worth noticing that, although not designed for that, the radiation monitor on-board LISA Pathfinder~\citep{Canizares11, armano2018GCR} allowed for the detection of the passage of large scale interplanetary structures such as high-speed solar wind streams and interplanetary counterparts of coronal mass ejections generating recurrent and non recurrent depressions of the galactic cosmic-ray flux~\citep{Armano18_GCR, Armano19_GCR}.

As we have previously discussed, in LISA Pathfinder the effect of the interplanetary magnetic field structures can not be easily distinguished in the absolute value of the measured magnetic field because the local magnetic field is largely dominated by the contribution of the spacecraft units. However, as we show in the following, the effect of these structures can have an impact in the spectra of fluctuations in the low frequency band, i.e. below the Hz. 

In order to trace the variability of the spectrum
we took a closer look at the amplitude of the spectra in two different frequency regions, 
namely $\rm 20 \ \mu Hz$ $< f_{LF} <$ $\rm 50 \ \mu Hz$ and  $\rm 20 \ mHz$ $< f_{HF} <$ $\rm 40 \ mHz$. 
We selected these two frequency regions because, as discussed above, 
magnetic field fluctuations come from a different physical origin, i.e.
interplanetary magnetic field and magnetometer read-out electronics, respectively.
For each of these frequency windows we compute the power spectral density.
We selected windows of 16 hours to compute each bin. 
In order to avoid segments containing magnetic experiments or events as the ones reported
in Tables \ref{tbl.magnetic_events} \& \ref{tbl.magnetic_injections}, we apply a mask to the data. To do so we take as a figure 
of merit the amplitude spectral density in 
the range $\rm 1 \ mHz < \ \it f < \rm 10\, mHz$. Those segments 
where this figure of merit is exceed by five sigma are discarded from our analysis.
Following this criterion, we exclude 9 segments out of 300
This analysis allows a generic description of the statistical behaviour of the fluctuations 
without any previous assumption on its stationarity. 
Figure~\ref{fig.power} shows the results for both frequency ranges.
In agreement with our previous analysis, the amplitude of the fluctuations 
for the low-frequency bin is coherent and follows 
the same statistical distribution for all four magnetometers.  
Bins are statistically distributed with similar median values, namely 
\begin{eqnarray}
\rm \left. \widetilde{S}^{1/2}_{B_z, PY} \right|_{\it f_{LF}} & = & \rm 177_{-53}^{+80} \, nT \ Hz^{-1/2},   \nonumber \\
\rm \left. \widetilde{S}^{1/2}_{B_z, MY} \right|_{\it f_{LF}}  & = & \rm 182_{-60}^{+81} \, nT \ Hz^{-1/2},   \nonumber\\
\rm \left. \widetilde{S}^{1/2}_{B_z, PX} \right|_{\it f_{LF}}  & = & \rm 191_{-62}^{+101}  \, nT \ Hz^{-1/2},   \nonumber\\
\rm \left. \widetilde{S}^{1/2}_{B_z, MX} \right|_{\it f_{LF}}  & = & \rm 189_{-64}^{+102}  \, nT \ Hz^{-1/2}.  \nonumber
\end{eqnarray}
These values are based on the 16th, 50th and 84th percentiles of the histogram. Also, in this case we can derive a common mean value for the fluctuations in this frequency range of 
$\rm 207 \pm 6 \ nT \ Hz^{-1/2}$. On the other hand, the  noise power in the higher frequency bins show a different amplitude spectral density for each magnetometer. The median values in this case are:

\begin{eqnarray}
\rm \left. \widetilde{S}^{1/2}_{B_z, PY} \right|_{\it f_{HF}} & = & \rm 7.4_{-0.3}^{+0.8} \, nT \ Hz^{-1/2},  \nonumber \\
\rm \left. \widetilde{S}^{1/2}_{B_z, MY} \right|_{\it f_{HF}}  & = & \rm 12.3_{-0.7}^{+2} \, nT \ Hz^{-1/2},   \nonumber\\
\rm \left. \widetilde{S}^{1/2}_{B_z, PX} \right|_{\it f_{HF}}  & = & \rm 10.6_{-0.6}^{+2}  \, nT \ Hz^{-1/2},   \nonumber\\
\rm \left. \widetilde{S}^{1/2}_{B_z, MX} \right|_{\it f_{HF}}  & = & \rm 9.0_{-0.6}^{+0.8}  \, nT \ Hz^{-1/2}.   \nonumber
\end{eqnarray}
In the latter we observe that the distribution of the median values are narrower and not overlapping between them. Both behaviours are clearly distinguished in the histograms of Fig.~\ref{fig.power}, which characterises the variability of the magnetic field fluctuations on-board for the two frequency regimes that we have previously identified.

\begin{figure}
\begin{center}
\includegraphics[width=0.5\textwidth]{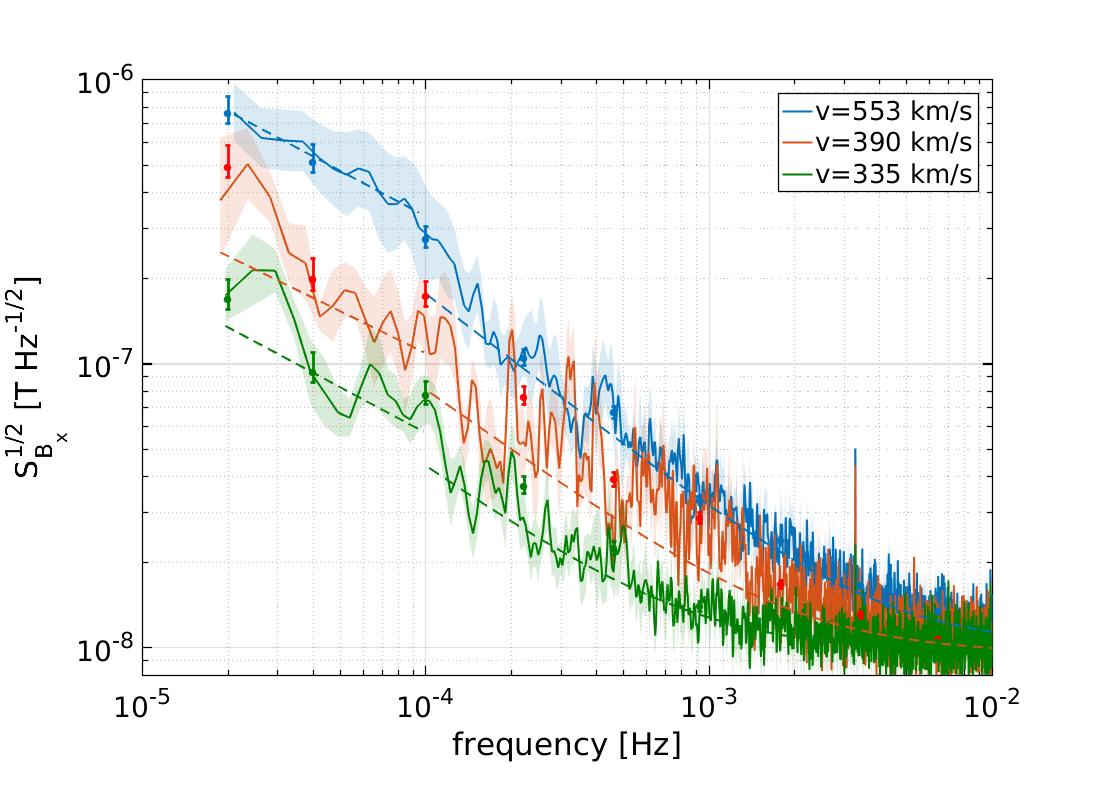}
\caption{Amplitude spectral density of the X-component of the magnetic field for 3 periods of different solar wind speed. The data to compute each spectrum has been obtained combining the data from the four magnetometers. The values of the fits to the data (dashed lines) are reported in Table~\ref{tab.fitValues}.\label{fig.SW}}
\end{center}
\end{figure}

In the framework of future gravitational-wave detectors in space, the variability in the spectra of magnetic field fluctuations is particularly relevant in the low frequency regime. There are two main reasons for that: first, the mHz band is the main objective of a gravitational-wave detector in space, since these observatories are designed to study the gravitational universe in this frequency band. Second, at the same time, the low frequency fluctuations will be precisely the main contribution to the magnetic induced force noise in the free falling test masses. According to \cite{Armano16}, magnetic induced forces could contribute to a maximum of 3 $\rm fm \ s^{-2} \ Hz^{-1/2}$ of the measured 12 $\rm fm \ s^{-2} \ Hz^{-1/2}$ at 0.1 mHz of the differential acceleration measured on LISA Pathfinder test masses. Although the precise determination of the magnetic noise contribution is still pending, the current estimate is expected to be of the order of other contributions such as the charging noise, with an expected contribution of 1 $\rm fm \ s^{-2} \ Hz^{-1/2}$~\citep{Armano2017} or the actuation noise, which is expected to be the dominant contribution with an expected value of 4.5 $\rm fm \ s^{-2} \ Hz^{-1/2}$~\citep{Armano16} at 0.1 mHz. For that reason it is worth characterising further the variability in this frequency regime to provide information for future space-borne observatories. As we show in the following, the fluctuations of the magnetic field in the sub-mHz band are deeply connected to the dynamics of the interplanetary plasma.

Fluctuations in the intensity of the interplanetary magnetic field can be associated with a wide variety of phenomena~\citep{Bruno2013}. However, particularly in the inertial range, the amplitude of the magnetic fluctuations is strictly related to the their Alfv\'enic nature. The solar wind is highly structured in high and low-speed streams, which carry different types of fluctuations. While fast wind is characterised by large-amplitude Alfv\'enic fluctuations, the slow wind generally advects less Alfv\'enic fluctuations characterised by a smaller amplitude --with the important exception of the Alfv\'enic slow wind, see~\citet{damicis2015}. This means that, moving from high- to low-speed regions, the power level of the magnetic fluctuations within the inertial range progressively decreases, though keeping the typical $f^{-5/3}$ Kolmogorov scaling. As a matter of fact, solar wind turbulence may be thought as superposition of a magnetic field background spectrum, common to both fast and slow flows ~\citep{Bruno2017}, and a turbulent large-amplitude Alfv\'enic spectrum, characteristic of the fast solar wind.



In order to study the impact of the solar wind speed in our measurements, 
we selected three segments representing stable periods of solar wind speed.
These periods had to be long enough to allow an estimate of amplitude spectral 
density down to $\rm 20\,\mu Hz$. 
The selected time spans correspond to Jul. 7th-14th 2016,  May 28th- Jun. 1st 2017 and Feb. 11th-16th 2017 when, 
according to measurements recorded by ACE, the solar wind had a mean velocity 
of $\rm 553 \pm 47\, km \ s^{-1}$, $\rm 390 \pm 53\, km \ s^{-1}$ and $\rm 335 \pm 35\, km \ s^{-1}$, respectively.
For each of these segments we evaluated the amplitude spectral density of the magnetic field as measured by LISA Pathfinder magnetometers. 
Although other authors have already studied this phenomena~\citep{Bruno2017}, we extended the characterisation to the sub-mHz regime, which is the region of greatest interest for LISA and future gravitational-wave detectors. 

Results are shown in Figure~\ref{fig.SW}, where we can distinguish an increase in the power of the low-frequency fluctuations that correlates with the increase of solar wind velocity. Indeed, we observe that fluctuations at $\rm 20\,\mu Hz$ vary from $\rm 170^{+30}_{-10} \,nT \ Hz^{-1/2}$ for a slow wind situation (typical velocities $\rm v \simeq 300 \ km \ s^{-1}$) to $\rm 750^{+100}_{-50}  \,nT \ Hz^{-1/2}$ when we consider a situation of high-speed wind (typical velocities $\rm v \simeq 500 \ km \ s^{-1}$). These two scenarios represent a deviation of 18\% and 362\%, respectively, with respect to the mean value that we have previously derived for the complete time series, $\rm 207 \pm 6 \,nT \ Hz^{-1/2}$.
Although other phenomena may also contribute to the variability of the sub-mHz fluctuations of the spectra, we consider this correlation with the solar wind as one of the physical mechanisms behind the statistical distribution of the fluctuations in the sub-mHz band that we obtained in Fig.~\ref{fig.power}. It is thus a dependence that future space-borne gravitational-wave detectors will need to take into account, since as we mentioned in section \ref{sec.magnetic_forces}, the coupling of the magnetic field fluctuations with the magnetic moment of the test masses could produce spurious forces that a gravitational-wave detector would sense as a change in the acceleration noise, and therefore, possibly mistaking it as a gravitational-wave signal .  

In Table \ref{tab.fitValues} we characterise the variation of the shape of the magnetic field fluctuations in terms of the solar wind speed. We fitted the spectra of the magnetic field fluctuations to two power laws, one between 20--100 $\mu$Hz and another one from 100 $\mu$Hz to 10 mHz. We have made this differentiation because, as previously explained, different mechanisms act at different time scales, resulting in different spectral indexes for these power laws. The fits were performed by means of a chi-squared minimization, non-linear fit using derivative-free method. For the sake of completeness, in Table \ref{tab.fitValues} we also show the results of the analysis for the Y and Z axis, which are not shown in the figure.

It is interesting to notice that, according to our parametrisation, the main impact of the wind speed is in the parameter B which accounts for the power of the fluctuations in the frequency band described by the power law. 
The values for the spectral index C are around 1.5, smaller than the value expected, which is 1.65, for the inertial range of interplanetary magnetic field fluctuations. This small difference may come from the fact that in the LPF case, the read-out noise of the magnetometers starts to flatten the spectrum around 1 mHz, and this may result in a slightly less steep curve when we perform the fit. Moreover, the inertial range of interplanetary magnetic field fluctuations starts between 0.1--1 mHz, while the magnetometers read-out noise starts to be dominant around 1 mHz. Therefore, we can not see a large and clear portion of the inertial range part of the interplanetary magnetic field spectrum.
With regard to the parameter A, the comparison between the different wind velocity regimes do not show any significant variation, as we expected given that this parameter describes the noise floor of the instrument, which is dominated by the electronics read-out contribution.
When we analyze the parameters corresponding to the lower frequency part, we find that the spectral index E values are around 1, which are in good agreement with the $f^{-1}$ behaviour of the interplanetary magnetic field fluctuations at this frequency range -- even though the errors are high because we do not have many points for the fit. Finally, the values obtained for the parameter D are not very well fitted, probably because of the small amount of data points available in this frequency range. 

\begin{table}
	\centering
	\caption{Values of the fits of the dashed lines shown in Fig.~\ref{fig.SW} to a function $A+B\, 2\pi f^{-C}$ in the range between $\rm 100 \ \mu Hz-10 \ mHz$ and to a function $D\, 2\pi f^{-E}$ in the range between $\rm 20 \ \mu Hz-100 \ \mu Hz$.}
	\label{tab.fitValues}
	\begin{tabular}{lccc} 
		\hline
		\multicolumn{4}{c}{\hspace*{2.5cm} Solar wind speed [km/s]}\\
		Parameter & $553 \pm 47$ & $390 \pm 53$ & $335 \pm 35$ \\	
		\hline
	    \multicolumn{4}{c}{$B_x$}     \\  
       \hline
		A $ (10^{-16} \times [\rm T^2/Hz])$ & $1.02 \pm 0.03$ & $0.90 \pm 0.02$ & $0.91 \pm 0.02$ \\
		B $ (10^{-19} \times [\rm T^2/Hz])$ & $3.8  \pm 0.6 $ & $1.8  \pm  0.4$ & $0.6  \pm  0.2$ \\
		C      [ ]                          & $1.53 \pm 0.03$ & $1.42 \pm 0.04$ & $1.41 \pm 0.06$ \\
		D $ (10^{-17} \times [\rm T^2/Hz])$ & $4    \pm   10$ & $1    \pm   30$ & $0.1  \pm  0.4$ \\
		E      [ ]                          & $1.1  \pm  0.3$ & $1.0  \pm  0.5$ & $1.0  \pm  0.4$ \\
       \hline
       	    \multicolumn{4}{c}{$B_y$}     \\  
       \hline
		A $ (10^{-16} \times [\rm T^2/Hz])$ & $1.11 \pm 0.03$ & $0.86 \pm 0.02$ & $0.94 \pm 0.02$ \\
		B $ (10^{-19} \times [\rm T^2/Hz])$ & $2.8  \pm 0.4 $ & $1.9  \pm  0.6$ & $0.3  \pm  0.1$ \\
		C [ ]                               & $1.58 \pm 0.03$ & $1.34 \pm 0.06$ & $1.53 \pm 0.07$ \\
		D $ (10^{-17} \times [\rm T^2/Hz])$ & $6    \pm   20$ & $0.001\pm0.006$ & $2    \pm   10$ \\
		E [ ]                               & $1.0  \pm  0.3$ & $1.7  \pm  0.4$ & $0.8  \pm  0.5$ \\
       \hline
              	    \multicolumn{4}{c}{$B_z$}     \\  
       \hline
		A $ (10^{-16} \times [\rm T^2/Hz])$ & $1.09 \pm 0.02$ & $0.91 \pm 0.02$ & $0.96 \pm 0.01$ \\
		B $ (10^{-19} \times [\rm T^2/Hz])$ & $1.2  \pm  0.2$ & $1.0  \pm  0.3$ & $0.22 \pm 0.08$ \\
		C [ ]                               & $1.62 \pm 0.03$ & $1.46 \pm 0.05$ & $1.60 \pm 0.06$ \\
		D $ (10^{-17} \times [\rm T^2/Hz])$ & $0.3  \pm  0.9$ & $10   \pm   20$ & $0.000\pm0.001$ \\
		E [ ]                               & $1.3  \pm  0.3$ & $0.6  \pm  0.1$ & $1.8  \pm  0.4$ \\
       \hline
       	\end{tabular}
\end{table}


\begin{figure}
\begin{center}
\includegraphics[width=0.5\textwidth]{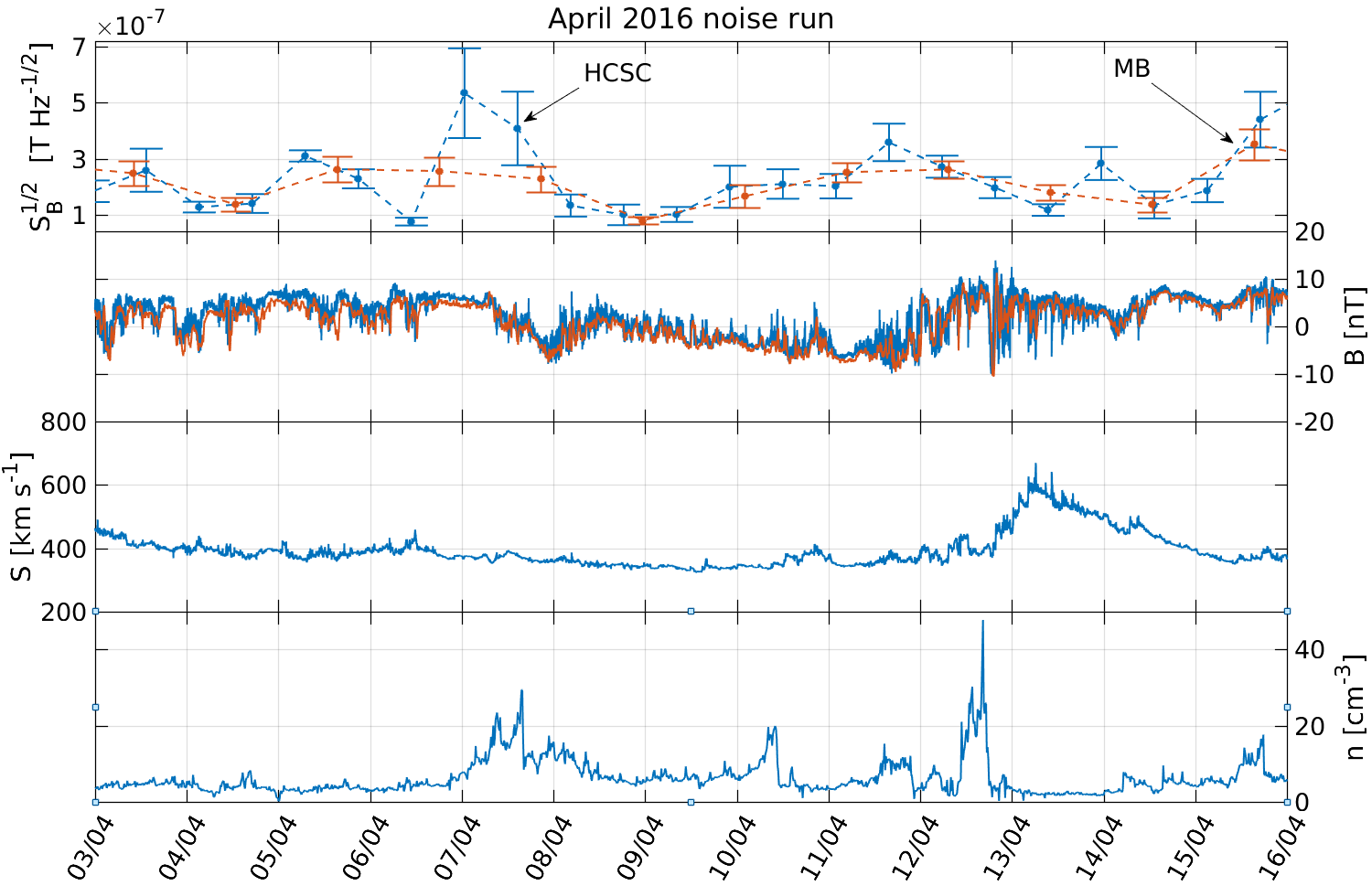}
~\\
\includegraphics[width=0.5\textwidth]{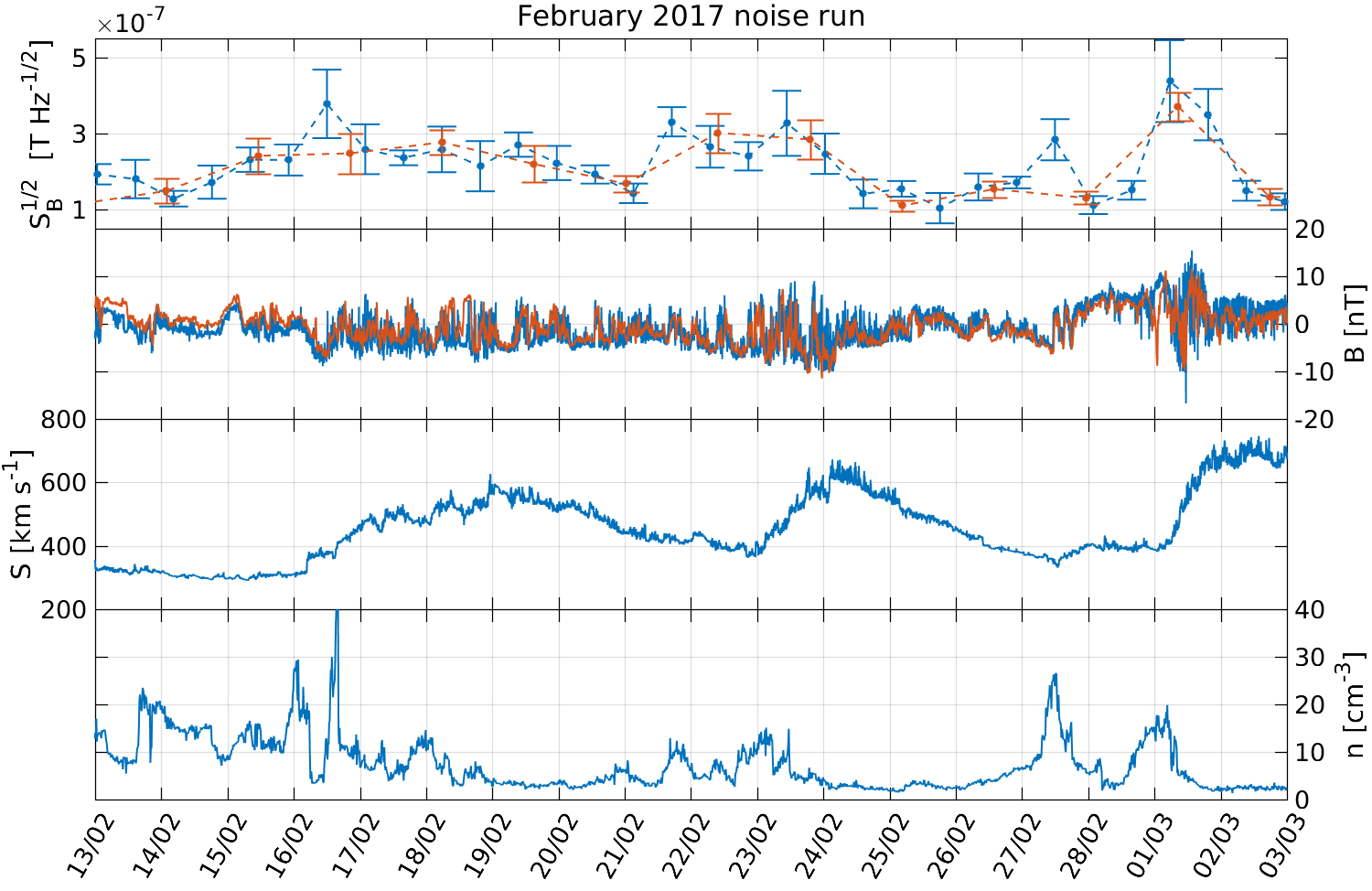}
\caption{\emph{Top:} Evolution during April 2016 noise run in LISA Pathfinder (dates indicated in figure) from (top to bottom):
\emph{-Upper Top:} Noise power in the 20--50 $\mu$Hz frequency bin for an average of the three components of LPF PY magnetometer (red) and ACE magnetometer (blue). The events listed are taken from~\protect\citet{Armano19_GCR} and noted in the text.
\emph{-Middle top:} Magnetic field $B_x$ component in GSE coordinates measured by ACE (blue), and LPF PY magnetometer $B_z$ component (red) with a lowpass at 1 mHz and with its mean value substracted.
\emph{-Middle bottom:} Solar wind velocity in GSE coordinates as measured by ACE.
\emph{-Lower bottom:} Proton density measured by ACE.
\emph{Bottom:} Same as top figure but for the February 2017 noise run performed in LISA Pathfinder (dates indicated in figure.)\label{fig.2plots}}
\end{center}
\end{figure}
 
To finalise the study of the non-stationarity of the magnetic field fluctuations, we took a closer look at the impact of this contribution during the noise performance runs. Analogously as it is done with on-ground gravitational-wave detectors during science runs, in LISA Pathfinder these runs were periods where the instrument was configured in its more sensitive configuration and kept in data acquisition mode without introducing any calibration signal. For LISA Pathfinder, these noise performance runs took typically 5 to 10 days although it is worth stressing that future space-borne gravitational wave detectors, such as LISA, aim to spend weeks or months in this scientific acquisition mode. 

In Figure \ref{fig.2plots} we analysed the non-stationarities of the low-frequency part of the magnetic field spectrum during the two LISA Pathfinder noise runs that were published in~\citet{Armano16} (April 2016 noise run) and~\citet{Armano18} (February 2017 noise run).
Following the same approach as we have previously shown, we computed the amplitude spectral density of the magnetic field in the frequency span that goes from 20--50 $\mu$Hz.
In the top panels of Figure \ref{fig.2plots} we show the evolution of this value over the duration of each LPF noise run for LPF and ACE satellites. 
We compare this with the evolution of the time series of the magnetic field itself $B_x$ component in Global Solar Ecliptic (GSE) coordinates (second panel) for ACE. In the same panel, we also show the magnetic field $B_z$ component as measured in LISA Pathfinder during the same period. The reason to compare LPF's $B_z$ in LPF reference frame with ACE's $B_x$ in GSE coordinates is that LPF's Z-axis points always towards the solar panel, which is always pointing towards the Sun. Thus, LPF positive Z-axis is roughly equivalent to positive X-axis in GSE coordinates.  
To do this comparison, LISA Pathfinder data has been low-pass filtered at 1 mHz in order remove the higher frequency noise coming from electronics and the mean substracted in order to exclude the DC magnetic field coming from the spacecraft components. As we previously showed when analysing the components in the spectra, the main features of the long term evolution are driven by the interplanetary component.  
As we saw, the phenomena associated with the solar wind requires several parameters that detail the characteristics of the interplanetary plasma. In order to provide a comprehensive view of this phenomena during these two particular noise runs we show as well the solar wind speed (third panel) and the proton density (fourth panel), as measured by ACE.

The results show that, during the February 2017 noise run, there are roughly three clear increases in the ASD of the low frequency bin of the magnetic field. These three peaks seem to be caused by high-speed streams in the solar wind speed, i.e, sudden increases in the solar wind speed, a type of event that also carries a decrease in the proton density, as we can also clearly observe in the lowest panel.
Regarding the April 2016 noise run, we can see another high-speed stream around April 13th that causes another small peak in the ASD of the low frequency bin of the magnetic field, associated with the corresponding decrease in proton density. Apart from that, there are two events that cause another increase in the time-series of the first panel. According to~\citet{Armano19_GCR}, these events would be associated to a Heliospheric current Sheet Crossing (HCSC) and a Magnetic Barrier (MB). We refer the reader to the aforementioned reference for more details on these events.

%% file: include/conclusions.tex
\section{Conclusions}
\label{sec.conclusions}

In this work we provide for the first time a description of the magnetic field on-board a gravitational-wave detector technology demonstrator in space. Our characterisation is tailored to address challenges created by the magnetic field that space-borne gravitational-wave observatories will face when they become operative. 

LISA Pathfinder included four magnetometers as part of its magnetic diagnostic subsystem. They were placed inside the thermal shield surrounding the optical bench and the vacuum chambers containing the test masses. Due to the magnetic content of the fluxgate core, the magnetometers had to be placed far enough from the test masses to prevent this active magnetic core to induce magnetic foces on the free-falling test masses. The major drawback of this configuration is the lack of resolution to measure the magnetic field or the magnetic field gradient at the location of the test masses. 


The DC magnetic field measured on-board is completely dominated by the contribution from the electronics of the spacecraft units. Among them, the thruster systems were a major contributor, both the cold gas high pressure latch valves (the ones used by ESA) and the colloidal thrusters (the ones operated by NASA). 
Cold gas thrusters or, more precisely, some permanent magnets in the cold gas thruster subsystem,  contribute with roughly the 80\% of the measured magnetic field. Although a strong contribution, this one remains constant throughout the mission -- partially thanks to the high thermal stability reached on-board~\citep{Armano19_Temp} -- which is key for a mission as LISA with strong requirements on any potential source of fluctuations.  
This is not the case for the colloidal thrusters, where we observed a persistent slow drift of around 150\, nT for the measurement of two magnetometers close to the test masses. This 150 nT drift took place over around 150 days, which corresponds to the first period of DRS operations -- see Fig. \ref{fig.timeseries} to check the different periods. 
We attribute this effect to the displacement of a stainless steel bellow inside the thruster that keeps the propellant at a constant pressure. Although not a desirable effect for future gravitational wave detectors in space, this slow change in the local magnetic field should not be present in future missions, since it could impact the main measurement of a mission like LISA. 

We took special care on the analysis of the fluctuations of the magnetic field on-board since these are a key component of the test mass force noise apportioning.  
We report how, on the one hand, the fluctuations of the magnetic field on-board are dominated at frequencies above the mHz by the contribution of the read-out electronics. In this frequency regime we identified and characterised an excess noise when compared to the design curve of our electronics. Although unexpected, this does not represent a major problem since the frequencies in which the magnetic noise may have an important impact are below the mHz.
On the other hand, below the mHz, the fluctuations are dominated by the contribution from the interplanetary magnetic field. Several indications point toward an interplanetary origin of the low-frequency fluctuations of the magnetic field. First, all magnetometers show coherent fluctuations in this frequency regime and, second, the densities of the magnetic field components measured by our four magnetometers match those measured by dedicated space weather missions ACE and WIND, which are also orbiting around L1.

Finally, we evaluated the non-stationary component affecting the very low-frequency regime of the magnetic field fluctuations. Due to its dependence with the solar wind, the low-frequency fluctuations show a large variability associated with changes in the interplanetary plasma. We tracked the amplitude spectral density in the low end of the LISA measuring band, i.e. at $\rm 20 \mu Hz$, during the whole duration of the mission. In this frequency regime, the magnetic field fluctuation on-board has a typical mean value of $\rm 207 \pm 6 \,nT \ Hz^{-1/2}$, although it shows an important variability with a wide statistical distribution of its values. 
Following previous studies, we show how this variability is tightly associated with a variety of phenomena associated with the dynamics of the interplanetary plasma. 
We have described and characterised how quantities describing the solar wind, as for example the plasma velocity, can be used to parametrise the variability of the low-frequency fluctuations. In the case of the solar wind velocity, we saw how variations in the range of $300-500$ km s$^{-1}$ are related to variations in the amplitude spectral density in the range $\rm 20-50 \mu Hz$ of around $\rm 170-750 \,nT \ Hz^{-1/2}$. 
We want to stress that the variability of the spectra in this low frequency regime is particularly relevant for LISA since these are frequencies in the measurement band of LISA and, also, where the magnetic fluctuations are important to the noise budget, as they are proportional to induced force noise in the test mass. 
Hence, variabilities of the order of 300\% in the spectra of magnetic field fluctuations (as the ones reported here) have to be taken into account in the design phase in order to prevent that this non-stationary behaviour impacts the performance of the future observatory. 

Having an instrument performance curve which is independent of the changes of the environment has been, historically, one of the main efforts of the gravitational wave community.
By using the data of  a pioneer mission (as LISA Pathfinder was) 
we expect to contribute to this decade long effort but now putting the focus on the interplanetary environment. We consider that our description and analysis on the magnetic field on-board and its associated potential sources of fluctuations will help the design of future long-term gravitational-wave observatories in space, like LISA will be.

%% file: include/acknowledgments.tex
\section*{Acknowledgments}

This work has been made possible by the LISA Pathfinder mission, 
which is part of the space-science program of the European Space Agency.
The French contribution has been supported by CNES (Accord Specific de 
projet CNES 1316634/CNRS 103747), the CNRS, 
the Observatoire de Paris and the University Paris-Diderot. 
E. P. and H. I. would also like to acknowledge the financial support of the UnivEarthS Labex program at Sorbonne Paris Cité (ANR-10-LABX-0023 and ANR-11- IDEX-0005-02).
The Albert-Einstein-Institut acknowledges the 
support of the German Space Agency, DLR. 
The work is supported by the Federal Ministry for Economic Affairs and Energy based on a resolution of the German Bundestag (FKZ 50OQ0501 and FKZ 50OQ1601).
The Italian contribution has been supported by 
Agenzia Spaziale Italiana and Instituto Nazionale di Fisica Nucleare. 
The Spanish contribution has been supported by Contracts No. AYA2010-15709 (MICINN), 
No. ESP2013-47637-P, and No. ESP2015-67234-P (MINECO). 
M. N. acknowledges support from Fundaci\'on General 
CSIC (Programa ComFuturo). F. R. acknowledges support from a Formaci\'on
de Personal Investigador (MINECO) contract.
The Swiss contribution acknowledges the support of 
the Swiss Space Office (SSO) via the PRODEX Programme 
of ESA. L. F. acknowledges the support of the Swiss National
Science Foundation.
The UK groups wish to acknowledge support from the
United Kingdom Space Agency (UKSA), the University of Glasgow, 
the University of Birmingham, Imperial College, 
and the Scottish Universities Physics Alliance (SUPA).
J.I.T. and J.S. acknowledge the support of the U.S. 
National Aeronautics and Space Administration (NASA).

%% file: include/appendix.tex
\section{Magnetic field absolute values}
\label{sec.DCvalues} 

As previously described in the text, the mean values of the magnetic field measured onboard were mostly constant through operations.
It was only due to some configuration changes in the satellite or due to induced magnetic fields during in-flight experiments that we did observe variations in the mean value of the magnetic field. 
In order to provide a more quantitative description, we summarise in Table~\ref{tbl.DCvalues}  the mean values of the three components of the magnetic field measured onboard for the four magnetometers. We repeated the same analysis for the four different phases that we already distinguished during our analysis, i.e. LTP and DRS nominal and extended operations. 
For each segment we also provide an estimate of the gradient of the magnetic field across each couple of magnetometers in the X and Y direction. 
The dates for each subset were selected trying to maximize the amount of data for that segment and trying to avoid any major changes (no experiments, glitches, etc). 	LTP nominal goes from March 01 2016 to April 27 2016. LTP extended goes from December 07 2016 to January 13 2017 and from January 15 2017 to March 14 2017. The 2 days gap is to get rid of the short DRS period in between. DRS nominal goes from June 27 2016 to December 7 2016. DRS extended goes from March 19 2017 to April 28 2017. Errors are computed as $\sigma/\sqrt{N}$.\\

\noindent \textit{Author affiliations}\\

\noindent$^{a}$ European Space Technology Centre, European Space Agency, 
Keplerlaan 1, 2200 AG Noordwijk, The Netherlands \\
 $^{b}$  Albert-Einstein-Institut, Max-Planck-Institut f\"ur Gravitationsphysik und Leibniz Universit\"at Hannover,
Callinstra{\ss}e 38, 30167 Hannover, Germany \\
 $^{c}$  APC, Univ Paris Diderot, CNRS/IN2P3, CEA/lrfu, Obs de Paris, Sorbonne Paris Cit\'e, France \\
 $^{d}$  Department of Industrial Engineering, University of Trent o, via Sommarive 9, 38123 Trento, 
and Trento Institute for Fundamental Physics and Application / INFN \\
 $^{e}$  Dipartimento di Fisica, Universit\`a di Trento and Trento Institute for 
Fundamental Physics and Application / INFN, 38123 Povo, Trento, Italy \\
 $^{f}$  Istituto di Fotonica e Nanotecnologie, CNR-Fondazione Bruno Kessler, I-38123 Povo, Trento, Italy \\
 $^{g}$  DISPEA, Universit\`a di Urbino ``Carlo Bo'', Via S. Chiara, 27 61029 Urbino/INFN, Italy \\
 $^{h}$  The School of Physics and Astronomy, University of
Birmingham, Birmingham, UK \\ $^{i}$  European Space Astronomy Centre, European Space Agency, Villanueva de la
Ca\~{n}ada, 28692 Madrid, Spain \\
 $^{j}$  Institut f\"ur Geophysik, ETH Z\"urich, Sonneggstrasse 5, CH-8092, Z\"urich, Switzerland \\
 $^{k}$  The UK Astronomy Technology Centre, Royal Observatory, Edinburgh, Blackford Hill, Edinburgh, EH9 3HJ, UK \\
 $^{l}$  Institut de Ci\`encies de l'Espai (CSIC-IEEC), Campus UAB, Carrer de Can Magrans s/n, 08193 Cerdanyola del Vall\`es, Spain \\
 $^{m}$  European Space Operations Centre, European Space Agency, 64293 Darmstadt, Germany \\
 $^{n}$  High Energy Physics Group, Physics Department, Imperial College London, Blackett Laboratory, Prince Consort Road, London, SW7 2BW, UK \\
 $^{o}$  Department of Mechanical and Aerospace Engineering, MAE-A, P.O. Box 116250, University of Florida, Gainesville, Florida 32611, USA \\
 $^{p}$  Physik Institut, Universit\"at Z\"urich, Winterthurerstrasse 190, CH-8057 Z\"urich, Switzerland \\
 $^{q}$  SUPA, Institute for Gravitational Research, School of Physics and Astronomy, University of Glasgow, Glasgow, G12 8QQ, UK \\
 $^{r}$  Department d'Enginyeria Electr\`onica, Universitat Polit\`ecnica de Catalunya,  08034 Barcelona, Spain \\
 $^{t}$  Gravitational Astrophysics Lab, NASA Goddard Space Flight Center, 8800 Greenbelt Road, Greenbelt, MD 20771 USA \\
 $^{u}$  Airbus Defence and Space, Gunnels Wood Road, Stevenage, Hertfordshire, SG1 2AS, United Kingdom \\
 $^{v}$  National Institute for Astrophysics, Astrophysical Observatory of Torino, Via Osservatorio 20, 10025 Pino Torinese (TO), Italy

\clearpage

\begin{table}
\scalebox{0.80}{\parbox{1.3\linewidth}{%
	\centering
	\caption{Mean values of the DC magnetic field for different spacecraft configurations.}
	\label{tbl.DCvalues}
	\begin{tabular}{|lcc|cc|} 
		\hline
		Magnetometer &  \multicolumn{2}{c}{LTP Nominal} & \multicolumn{2}{|c|}{LTP Extended}  \\
		\hline
	                         & $(B_x, B_y, B_z)$ [nT]  & $ |\vec{B}|$ [nT]  &  $(B_x, B_y, B_z)$ [nT]  & $ |\vec{B}|$ [nT]    \\  
       \hline
		PX  & $(866.880 \pm 0.005, -908.001 \pm 0.005, 82.053 \pm 0.005)$ & $1258.046 \pm 0.009$ &  $(736.202 \pm 0.006, -896.577 \pm 0.005 , 100.497 \pm 0.006)$ & $1164.450 \pm 0.010$\\         
	    MX  & $(816.303 \pm 0.005, -457.364 \pm 0.006,  91.730 \pm 0.005)$ &  $940.184 \pm 0.009$ &$(696.128 \pm 0.005, -428.666 \pm 0.005 , 94.944 \pm 0.006)$ & $823.021 \pm 0.009$\\	            
		PY  & $(-111.894 \pm 0.005, 585.993 \pm 0.005, 384.374 \pm 0.005)$ & $709.684 \pm 0.009  $ & $(-131.575 \pm 0.005, 582.902 \pm 0.005 , 394.238 \pm 0.005)$& $715.898 \pm 0.009 $\\	           
	    MY  & $(-86.583 \pm 0.006, 1023.586 \pm 0.006, 527.776 \pm 0.005)$ & $1154.890 \pm 0.010  $ & $(-109.538 \pm 0.006,  1006.245 \pm 0.006 , 541.992 \pm 0.006)$& $1148.165 \pm 0.010     $\\	
       \hline
	                         & $(\partial_x B_x, \partial_x B_y, \partial_x B_z)$ [nT/m]  & $ |\partial_x \vec{B}|$ [nT/m]  &  $(\partial_x B_x, \partial_x B_y, \partial_x B_z)$ [nT/M]  & $ |\partial_x \vec{B}|$ [nT/m]   \\  
       \hline
		PX- MX  &   $(62.717 \pm 0.006, -611.033 \pm 0.007, -12.544 \pm 0.005)$ & $614.371 \pm 0.010$ &  $(49.694 \pm 0.006, -634.456 \pm 0.006 , 7.024 \pm 0.006)$ &  $636.438 \pm 0.010$\\
       \hline
	                         & $(\partial_y B_x, \partial_y B_y, \partial_y B_z)$ [nT/m]  & $ |\partial_y \vec{B}|$ [nT/m]  &  $(\partial_y B_x, \partial_y B_y, \partial_y B_z)$ [nT/M]  & $ |\partial_y \vec{B}|$ [nT/m]   \\  
       \hline
		PY- MY   &  $( -34.318 \pm 0.006, -542.507 \pm 0.005, -185.989 \pm 0.005)$ & $574.529 \pm 0.009$ &  $(-29.880 \pm 0.006, -524.843 \pm 0.006 , -191.632 \pm 0.006)$ & $559.532 \pm 0.010$ \\		
		\hline
		\hline
		Magnetometer &  \multicolumn{2}{c}{DRS Nominal} & \multicolumn{2}{|c|}{DRS Extended}  \\
		\hline
	                         & $(B_x, B_y, B_z)$ [nT]  & $ |\vec{B}|$ [nT]  &  $(B_x, B_y, B_z)$ [nT]  & $ |\vec{B}|$ [nT]    \\
       \hline
		PX  & $ 833.375 \pm 0.026, -901.685 \pm 0.004, 60.990 \pm 0.005)$ & $1229.337 \pm 0.026$ &  $(762.165 \pm 0.011, -886.156 \pm 0.007 , 79.828 \pm 0.009)$ & $1171.555 \pm 0.016$\\         
	    MX  & $(710.415 \pm 0.024, -437.246 \pm 0.006,  87.603 \pm 0.003)$ &  $838.778 \pm 0.025$ &$(634.580 \pm 0.010, -421.112 \pm 0.005 ,   94.165 \pm 0.006)$ & $767.394 \pm 0.013$\\	            
		PY  & $(-124.643 \pm 0.005, 581.454 \pm 0.003, 390.371 \pm 0.003)$ & $711.347 \pm 0.007  $ & $(-118.939 \pm 0.005, 583.369 \pm 0.005 , 391.387 \pm 0.006)$& $712.495 \pm 0.009 $\\	           
	   MY  & $(-102.913 \pm 0.006, 1064.324 \pm 0.006, 531.317 \pm 0.004)$ & $1194.016 \pm 0.008  $ & $(-102.051 \pm 0.008,  1052.539 \pm 0.011 , 551.385 \pm 0.008)$& $1192.593 \pm 0.016     $\\	
              
       \hline
	                         & $(\partial_x B_x, \partial_x B_y, \partial_x B_z)$ [nT/m]  & $ |\partial_x \vec{B}|$ [nT/m]  &  $(\partial_x B_x, \partial_x B_y, \partial_x B_z)$ [nT/m]  & $ |\partial_x \vec{B}|$ [nT/m]   \\  
       \hline
		PX- MX  &   $(151.504 \pm 0.011, -627.590 \pm 0.018, -34.241 \pm 0.007)$ & $646.5250 \pm 0.022$ &  $(158.210 \pm 0.015 , -630.569 \pm 0.006 , -18.599 \pm 0.008)$ &  $650.380 \pm 0.018$\\
       \hline
	                         & $(\partial_y B_x, \partial_y B_y, \partial_y B_z)$ [nT/m]  & $ |\partial_y \vec{B}|$ [nT/m]  &  $(\partial_y B_x, \partial_y B_y, \partial_y B_z)$ [nT/m]  & $ |\partial_y \vec{B}|$ [nT/m]   \\  
       \hline
		PY- MY   &  $(-29.619 \pm 0.005, -596.135 \pm 0.020, -182.260 \pm 0.006)$ & $624.078 \pm 0.021$ &  $(-22.898 \pm 0.010, -581.655 \pm 0.013 , -207.513 \pm 0.006)$ & $617.987 \pm 0.017$ \\		
		\hline
	\end{tabular}
	}}
\end{table}